\documentclass{aa}  

\usepackage{graphicx}
\usepackage{txfonts}
\usepackage{natbib,twoopt}
\bibpunct{(}{)}{;}{a}{}{,}             
\makeatletter
  \newcommandtwoopt{\citeads}[3][][]{\href{http://adsabs.harvard.edu/abs/#3}%
    {\def\hyper@linkstart##1##2{}%
     \let\hyper@linkend\@empty\citealp[#1][#2]{#3}}}
  \newcommandtwoopt{\citepads}[3][][]{\href{http://adsabs.harvard.edu/abs/#3}%
    {\def\hyper@linkstart##1##2{}%
     \let\hyper@linkend\@empty\citep[#1][#2]{#3}}}
  \newcommandtwoopt{\citetads}[3][][]{\href{http://adsabs.harvard.edu/abs/#3}%
    {\def\hyper@linkstart##1##2{}%
     \let\hyper@linkend\@empty\citet[#1][#2]{#3}}}
  \newcommandtwoopt{\citeyearads}[3][][]%
    {\href{http://adsabs.harvard.edu/abs/#3}
    {\def\hyper@linkstart##1##2{}%
     \let\hyper@linkend\@empty\citeyear[#1][#2]{#3}}}
\makeatother

\defcitealias{Cho16}{MIST}
\defcitealias{Hje87}{HW87}
\defcitealias{Pav15}{PI15}
\defcitealias{Che08}{CH08}
\defcitealias{Woo11}{WI11}
\defcitealias{Ge20a}{Ge+20}
\defcitealias{Zor11}{Zor11}

\usepackage[usenames,dvipsnames]{color}

\definecolor{pinegreen}{RGB}{1, 121, 111} 
\definecolor{violet}{RGB}{214,39,40}
\usepackage[breaklinks=true,colorlinks,urlcolor  = blue, linkcolor=violet,citecolor=pinegreen]{hyperref} 

\begin{document} 

\title{Coping with loss}
\subtitle{Stability of mass transfer from post-main-sequence donor stars}

   \author{     
            K. D. Temmink
            \inst{1}
            \fnmsep\thanks{e-mail: \href{mailto:Karel.Temmink@ru.nl}{Karel.Temmink@ru.nl} }
        \and
            O. R. Pols
            \inst{1}
        \and
            S. Justham
            \inst{2,3,4,5,6}
        \and
            A. G. Istrate
            \inst{1}
        \and
            S. Toonen
            \inst{2}
          }

   \institute{  
            Department of Astrophysics/IMAPP, Radboud University Nijmegen, P.O. Box 9010, 6500 GL    Nijmegen, The Netherlands
        \and
            Anton Pannekoek Institute for Astronomy, University of Amsterdam, 1090 GE Amsterdam,
            The Netherlands
        \and
            GRAPPA, University of Amsterdam, Science Park 904, 1098 XH Amsterdam, The Netherlands
        \and
            School of Astronomy \& Space Science, University of the Chinese Academy of Sciences, Beijing 100012, China
        \and 
            National Astronomical Observatories, Chinese Academy of Sciences, Beijing 100012, China
        \and 
            Max-Planck-Institut f\"{u}r Astrophysik, Karl-Schwarzschild-Stra{\ss}e 1, 85741 Garching, Germany
             }

   \date{Received xxx; accepted yyy}

 
  \abstract
   {The stability of mass transfer is critical in determining pathways towards various kinds of compact binaries, such as compact main-sequence  white-dwarf binaries, and transients, such as double white-dwarf mergers and luminous red novae. Despite its importance, very few systematic studies of the stability of mass transfer exist.} 
   {We study the behaviour of mass-losing donor stars in binary systems in a systematic way. We focus on identifying and understanding the parameter space for stable mass transfer in low- and intermediate-mass binaries with post-main-sequence donor stars as well as the properties of ultimately unstable binary systems at the onset of the instabilities.}
   {We employed the 1D stellar evolution code MESA to simulate the mass-transfer evolution of 1404 binary systems with donor-star masses between $1~M_\odot$ and $8~M_\odot$. We studied the behaviour of the binaries during mass transfer, without assuming that the donor star responds adiabatically to mass loss. We treated the accretor as a point mass, which we do not evolve, and assumed the mass transfer is conservative.}
   {%
   We considered several criteria to define when unstable mass transfer occurs. We find that the criterion that best predicts the onset of runaway mass transfer is based on the transition to an effectively adiabatic donor response to mass loss. Using this quasi-adiabatic criterion, we determine the location of the stability boundary to within a relative uncertainty of five per cent in the mass ratio at the onset of mass transfer. Defining this critical mass ratio ($q_{\rm qad}$) in terms of accretor mass over donor mass, we find that $q_{\rm qad}\sim0.25$ for stars with radiative envelopes that cross the Hertzsprung gap, while for convective giants $q_{\rm qad}$ decreases from $\sim 1$ at the base of the red giant branch to $\sim 0.1$ at the onset of thermal pulses on the asymptotic giant branch. Compared with recent similar studies, we find increased stability of mass transfer from convective giants. This is because an effectively adiabatic response of the donor star only occurs at a very high critical mass-transfer rate due to the short local thermal timescale in the outermost layers of a red giant. Furthermore, we find that for $q > q_{\rm qad}$ mass transfer is self-regulated, but that for evolved giants the resulting mass-transfer rates can be so high that the evolution becomes dynamical and/or the donor can overflow its outer lobe.}
   {Our results indicate that mass transfer is stable for a wider range of binary parameter space than typically assumed in rapid binary population synthesis. Moreover, we find a systematic dependence of the critical mass ratio on the donor star mass and radius, which may have significant consequences for predictions of post-mass-transfer populations.}


   \maketitle
%

\section{Introduction}
\label{sec:introduction}
A significant fraction of stars exists in multiple systems where the components will interact \citep[][]{Abt76,Abt83, Duq91, San12, Moe17}. As such, mass transfer is a crucial process in the evolution of many stellar systems \citep[e.g.][]{Mor60,Paczynski1971}, including for the formation of white-dwarf mergers \citep{Ibe84}, core-collapse supernova diversity \citep{Pos92}, and binary gravitational-wave sources \cite[e.g.][]{LIGO2016,LIGO2017,LISA22}.
Whether the onset of Roche-lobe overflow (RLOF) in a particular binary leads to stable or unstable mass transfer produces a qualitative difference in the future evolution, as well as in the properties of the final remnants. Where unstable mass transfer and the resulting common-envelope (CE) phase typically result in close binaries or a single, merged object \citep[e.g.][]{Pac76}, stable mass transfer tends to produce wider binaries \citep[e.g.][]{So97}. Different assumptions regarding mass-transfer stability therefore strongly affect predictions for, for example, double-white-dwarf (DWD) binaries and their associated transients, such as thermonuclear supernovae and gravitational-wave transients. These assumptions also affect transients linked to non-compact stars, such as luminous red novae, which are thought to be an observational manifestation of unstable mass transfer \citep{Tyl06,Tyl11,Iva13}. Despite its clear importance, only a few systematic studies of the stability of mass transfer, which we discuss in more detail below, have been performed over the years.

As is well known, the response of a star to mass loss depends on the structure of its envelope. The canonical
view \citep[e.g.][]{Web85} is as follows. Donor stars with radiative envelopes tend to shrink on their dynamical timescales in response to a sudden loss of mass. Therefore, mass transfer involving this type of donor star is expected to be relatively stable. If a donor star with a convective envelope loses mass, its radius tends to increase, unless the core-mass fraction exceeds 0.5. On the other hand, its Roche radius typically decreases if the donor star is more massive than the accretor. This means that such a donor star will overfill its Roche lobe by an ever-increasing amount, eventually leading to mass transfer on a dynamical timescale and the formation of a CE. Indeed, classical studies \citep[such as][]{Pac65,Pac69,Web85,Hje87} found that mass transfer from a red giant donor star to a less massive companion should be unstable for nearly the entire giant branch.

However, this classical picture relies on simplifying assumptions, such as the approximation of the stellar structure by simple (composite or condensed) polytropes, and the assumption that the donor star responds to the loss of mass fully adiabatically. It has long been known that this is a serious limitation \citep[see e.g.][]{Pac65}, and these results should be interpreted and applied with care. In fact, more recent theoretical studies \citep[see e.g.][]{Han02,Pos02,Woo11,Pas12,Pav15} found stable mass transfer for a wider range of mass ratios and orbital periods than the classical results mentioned above. These authors argued that the ability of the outer layers of giant stars to thermally readjust on dynamical timescales should significantly increase the stability of mass transfer compared to the classical results. Unfortunately, these studies typically focus on only a small part of the parameter space. An exception to this is the collection of recent work of \cite{Ge10,Ge15,Ge20a}, which does map the stability boundary systematically for a large set of masses, periods, and mass ratios. While the authors use a realistic equation of state to model the stellar structure, this set of studies still employs the adiabatic approximation mentioned above. 

Additionally, over the years, different physically motivated criteria have been proposed to identify unstable mass transfer in both detailed 1D simulations and rapid binary population synthesis (BPS) methods. Indeed, there are multiple physical reasons why mass transfer might become unstable \citep[see e.g.][]{Iva20}. These reasons include dynamical timescale evolution (which has become a common assumption, as in the references above) and the overflowing of an outer lobe (OL) in the Roche potential. As we explain in more detail in Sects. \ref{sec:criteria} and \ref{sec:results}, these criteria typically do not identify unstable mass transfer consistently; mass transfer can still be stable and self-regulating even when these criteria are met.

Not only the definition of reliable criteria for stable mass transfer poses a potential problem for the results mentioned above. As outlined in, for example, the discussion of \cite{Pos92}, there are several observed systems that are in tension with the predicted limits for stable mass transfer from polytropic adiabatic calculations. This includes systems that appear to have experienced mass transfer from donors on the red giant branch (RGB; case B) or asymptotic giant branch (AGB; case C) but also have relatively long orbital periods \citep[see e.g.][]{Egg89}. This suggests that this phase of mass transfer was likely stable, whereas classical results would have predicted otherwise. Additionally, \cite{Pos92} note that several massive systems are observed to be in such a phase of late case B or early case C mass transfer (e.g. VV Cep; \citealp{Wri77}, AZ Cas; \citealp{Cow77}, and KQ Pup; \citealp{Cow65}) but have apparently avoided the dynamical instability predicted by the classical results above.

Further populations of observed systems have long been known to indicate problems with classical, simplified results regarding the stability of mass transfer. For example, \cite{Tau99} and \cite{Pos02} found that the increased mass-transfer stability from their detailed stellar calculations, as compared to classical stability criteria, helps explain the properties of low-mass X-ray binaries and their descendants. Hot subdwarfs formed by stable mass transfer from a giant star onto a lower-mass companion also naturally suggest mass transfer that is more stable than predicted by the classical results (as noted and investigated by e.g. \citealt{Han02,Han03,Vos+2019,Vos+2020}). Stars that have become blue stragglers after stably accreting from a giant companion also motivate consideration of increased mass-transfer stability \citep[e.g.][]{Che08}.  \cite{Lei21} constructed multiple synthetic blue straggler populations using a range of different mass-transfer stability prescriptions. They find that none could reproduce the observed \textit{Gaia} sample adequately and stressed that a better understanding of the stability of mass transfer is needed to explain observations that point towards more stable mass transfer.  \cite{Woo12} were able to reproduce a subset of the observed DWD population by assuming the first phase of mass transfer is stable. Despite the fact that it has long been clear that assumptions made in rapid BPS methods tend to underestimate the stability of mass transfer, in particular the stability boundaries of \cite{Hje87} and \cite{Hur02}, they still remain widely in use today.

In this work we present a systematic study of mass-transfer stability for RLOF from donor stars with zero-age main-sequence (ZAMS) masses in the range $1~M_\odot~\leq M_{\rm d,ZAMS}~\leq~8~M_\odot$. We simulate the evolution of 1,404 binaries with the Modules for Experiments in Stellar Astrophysics (MESA) code \citep[see][]{Pax11, Pax13, Pax15, Pax18, Pax19} in order to obtain the boundary between stable and unstable mass transfer for stars that fill their Roche lobes as post-main-sequence (post-MS) stars, without assuming the donors respond to the loss of mass adiabatically.  In our simulations we only find such an adiabatic response when the mass-transfer rate exceeds a critical value, which is several orders of magnitude higher than the rate of mass loss occurring on the Kelvin-Helmholtz (KH) timescale of the donor (see Sect. \ref{sec:crit_th}) We introduce a criterion for stable mass transfer based on this transition to complete adiabatic evolution (our `quasi-adiabatic' criterion) and use it to identify the boundary between stable and unstable mass transfer. Furthermore, we also find alternative boundaries assuming two commonly used stability criteria, which are based on dynamical-timescale evolution and the overflowing of its OL by the donor star. We treat the accreting star as a point mass and therefore do not follow its evolution.

The contents of this work are organised as follows. In Sect.~\ref{sec:method} we describe our numerical method and assumptions. In Sect.~\ref{sec:criteria} we describe the three different physically motivated instability criteria we employ, and argue why we base our main results on the quasi-adiabatic criterion (Sect.~\ref{sec:crit_th}). We present our results in Sect.~\ref{sec:results_mdot}, where we focus on evolutionary aspects of the mass transfer, and Sect.~\ref{sec:results_qcrit}, where we focus on the critical mass ratios for stable mass transfer. We critically analyse the limitations of our simulations and results in Sect.~\ref{sec:discussion}, where we also compare our theoretical results to other results from the literature (Sect.~\ref{sec:discussion_comparison}) and to observations of post-mass transfer binaries (Sect.~\ref{sec:discussion_observations}). We end by summarising our conclusions in Sect.~\ref{sec:conclusion}.

\section{Numerical methods and physical assumptions}
\label{sec:method}

 We used MESA \citep{Pax11, Pax13, Pax15, Pax18, Pax19}, version \texttt{r12115}, to simulate the evolution of single stars with masses of 1.0, 1.5, 2.0, 2.5, 3.0, 4.0, 5.0, 6.0, 7.0, and 8.0$~M_\odot$, as well as to compute a grid of binary models. In this section we describe our most important assumptions for the single-star models, followed by a description of our binary evolution calculations. For more details regarding our MESA settings, we refer the reader to the MESA inlists (input files) used in this work\footnote{These are available at \href{https://zenodo.org/communities/mesa}{https://zenodo.org/communities/mesa}}. The computed single and binary star models are available on request, as are the other data used in this work \footnote{Additionally, the data in Tables \ref{tab:mdot_results} and \ref{tab:main_results} can be openly accessed in electronic form at the CDS via anonymous ftp to cdsarc.u-strasbg.fr (130.79.128.5) or via \href{http://cdsarc.u-strasbg.fr/viz-bin/qcat?J/A+A/}{http://cdsarc.u-strasbg.fr/viz-bin/qcat?J/A+A/} }.
 
 We adopt a slightly super-solar metallicity of $Z = 0.02$ \citep[see e.g.][]{Asp09, Mag22}. For the rest of our settings and assumptions, we generally follow the solar-calibrated assumptions described in \citet{Cho16}. We use the Ledoux criterion \citep{Led47} for determining whether or not a region inside the stars is stable against convection, and use a value for the mixing length parameter of $\alpha_{\rm MLT} = 1.82$ for convective regions. We also include the effects of semi-convection \citep[implemented as in][]{Lan83}, with $\alpha_{\rm semi} = 0.1$. We treat convective overshooting of stellar cores and burning shells as in the solar calibrated models of \citet{Cho16}, which were calibrated by matching models to the main-sequence turnoff in the open cluster M67.We employ MESA's default  \texttt{basic} nuclear network to model nuclear energy generation inside the donor stars, which is sufficient for our purposes and for the parameter space we consider. We also include mass loss through stellar winds: we assume a modest `Reimers' wind \citep{Rei75} with strength $\eta_R = 0.1$ on the RGB, and a `Bl\"ocker' wind \citep{Blo95} with strength $\eta_B = 0.2$ on the AGB.  We furthermore increase the resolution of the numerical mesh with respect to the default settings around the boundaries of convective zones, as well as in regions with strong H or He abundance gradients, which typically results in around 1,500 zones. We end our single-star simulations when the first thermal pulse (TP) occurs. 
 
\begin{figure}
\centering
\includegraphics[width=\hsize]{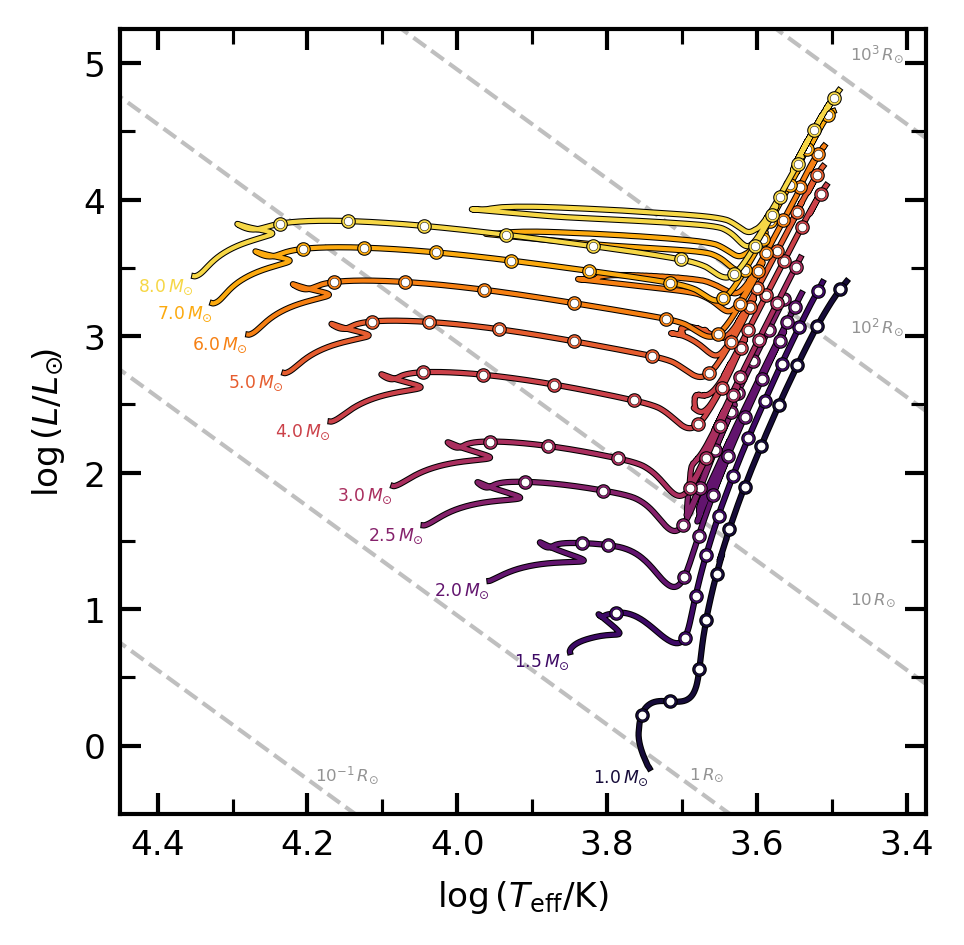}
  \caption{Single-star evolutionary tracks in the HRD for different stellar masses. Each colour corresponds to a given initial mass, as indicated to the lower left of each line. The coloured open circles indicate where we save single-star models for use in our binary star simulations. We have indicated lines of constant radius using grey dashed lines. Note that we have truncated these tracks to start at the ZAMS, and end at either the first TP or the maximum radius of the star, depending on which occurs first.}
     \label{fig:model_selection}
\end{figure}

We use the resulting single-star tracks to construct a grid of input models for our binary stellar simulations. For each track, we save models at pre-determined radii, which we then use as starting points in our MESA binary simulations. The points along the evolutionary tracks where models are saved are chosen by requiring a minimum radial resolution of $\Delta \log R_{\rm d} = 0.2$ across the entire track. We furthermore ensure that the smallest radius chosen for each evolutionary phase is larger than the largest stellar radius reached during the previous phase. For stars with masses $< 2~M_\odot$, we do not consider the AGB phase, as we find that these stars do not expand significantly beyond their maximum radius on the RGB before the first TP occurs. The above selection criteria result in about $11$ points in total for each donor mass. Figure \ref{fig:model_selection} shows our single-star evolutionary models in the Hertzsprung-Russell diagram (HRD), as well as the aforementioned points where we initialise mass transfer in binaries. 

For each of those saved models, we initialise 12 binary systems, with mass ratios ($q=M_{\rm a}/M_{\rm d}$; throughout this work, the subscripts `a' and `d' will refer to the accretor and donor, respectively) ranging from $0.1$ to $1.2$ in increments of $0.1$, for a total of 1404 binary star models. For each binary, we thus set $M_{\rm a} = q M_{\rm d}$, where $M_{\rm d}$ is the actual mass of the donor star (which is not necessarily equal to its initial mass) at each of the pre-determined points, including mass lost due to winds. We purposely include mass ratios $>1$ in our grids to ensure we always enclose the stability boundary within our grid points. Previous studies (see the references in Sect. \ref{sec:introduction}) indicate that for a sizeable part of our parameter space, the critical mass ratio (above which mass transfer is stable) could be above unity. We chose the initial orbit of the binary such that the donor star fills its Roche lobe (with equivalent-volume radius $R_{\rm L_1}$) at the pre-determined stellar radii (see above). To that end, we used the fitting formula from \cite{Egg83} (their Eq. 2) to calculate the size of the initial orbits, given the radii of the donor stars from our saved single-star models:
\begin{align}
\label{eq:rL1}
    R_{\rm L_1}/a \approx& \frac{0.49{q}^{-2/3}}{0.6{q}^{-2/3} + \ln(1 + {q}^{-1/3})},
\end{align}
where $a$ denotes the binary orbital separation.We note that we make the simplification that the initial eccentricity is equal to $0$.

We generally use the same settings and configurations for the donor star in the binary system as for the single-star simulations described above. Additionally, we use the `\texttt{Kolb}' scheme \citep{Kol90} to calculate the mass-transfer rate. As previously mentioned, we do not evolve the accreting star, and take it to be a point mass. This allows us to study the stability of the donor stars separately from the evolution of the accretor. Additionally, simulating accreting stars in binaries requires special attention to the proper treatment of many quantities, such as the specific entropy of the accreted material and the spin of the accretor, which is outside the scope of this work (but see Sect. \ref{sec:discussion}). 

We furthermore assume that all the material lost by the donor is accreted by its companion, that is, the mass transfer phases are fully conservative in terms of both mass and angular momentum. This assumption cannot be expected to hold in all or even most circumstances. The effect of non-conservative RLOF on the stability of mass transfer strongly depends on assumptions regarding the loss of angular momentum. For example, in the case where any material that is not accreted is assumed to leave the system with the specific angular momentum of the accretor (isotropic re-emission), it can be shown analytically that an increase in the fraction of mass leaving the system ($\beta$) results in a decrease of ${\rm d} \ln R_{\rm L_1}/{\rm d}\ln M_{\rm d}$ and ${\rm d}\ln a/{\rm d}\ln M_{\rm d}$ \citep[see e.g.][]{So97, Tau99, Tau96}, making mass transfer more stable. Hence, our results should provide an upper limit (in terms of the critical $M_{\rm a}/M_{\rm d}$) on the location of the boundary between stable and unstable mass transfer in the parameter space, that is, mass transfer is expected to be more stable than found here if mass were lost from the system under the assumption of isotropic re-emission (but see Sect. \ref{sec:discussion_limitations} for further discussion).

In this work, we do not model stellar rotation and thus do not include effects of spin-orbit coupling on the evolution of the binaries and the stability of mass transfer \citep[but see][for an extensive study]{Mis2020}. Hence, we also do not keep track of the possible occurrence of Darwin instabilities \citep{Dar1879}, which may become relevant for extreme mass ratios $M_{\rm a}/M_{\rm d} \la 0.1$ (but see Sect.~\ref{sec:discussion_limitations}).

We focus on the first phase of mass transfer and end our calculations when the donor star has shrunk well inside its Roche lobe, either after its envelope has mostly been exhausted or after it has started He burning in its core. As such, in this work we do not consider subsequent episodes of mass transfer that could occur when partially stripped giants re-expand to fill their Roche lobes after a core He-burning phase.

Having simulated the grid points mentioned above, we systematically decrease the uncertainty on the location of the stability boundary resulting from our main criterion (see Sect. \ref{sec:criteria}) by using a simple bisection method. We execute sufficient bisection iterations such that the relative uncertainty on the critical mass ratio is $5\%$ or less. For the other two criteria, we interpolate linearly in our grid points to estimate the critical mass ratios. Finally, we systematically verify that our results are converged by re-simulating a representative selection of stable and unstable systems with increased resolution in both time and mesh spacing. We discuss these tests in more detail in Appendix \ref{sec:convergence_study}.

\section{Criteria for unstable mass transfer}
\label{sec:criteria}
A classical approach \citep[e.g.][]{So97} for determining whether or not mass transfer is stable, is to compare the adiabatic response to mass loss of the donor star radius $\zeta_{\rm ad} \equiv {\rm d}\log R_{\rm d}/{\rm d}\log M_{\rm d}$ to the response of its Roche lobe ($\zeta_{L} \equiv {\rm d}\log R_{\rm L}/{\rm d}\log M_{\rm d}$). If the radius of the donor star persistently increases relative to its Roche lobe during the mass transfer, a runaway situation develops and the mass transfer is assumed to become unstable. Ultimately, a CE that engulfs both stars may form \citep{Pac76}. However, if the donor star can remain roughly equal in size or even shrink relative to its Roche lobe, the mass transfer is stable and self-regulating. This comes with the caveat that, even if at any point during the mass transfer $\zeta_{\rm ad} < \zeta_{\rm L}$ briefly, mass transfer does not necessarily become unstable, provided that the donor star recovers and shrinks relative to its Roche lobe again. As such, one would need to consider the entire evolution of the mass transfer, and it is not always possible or practical to accurately predict the outcome of the mass transfer at the onset using the $\zeta$ method. Well-known examples of this are so-called delayed dynamical instabilities \citep[see e.g.][]{Hje87,Han06,Pav15, Ge20a}, where mass transfer is initially stable (according to the $\zeta$ method), but only turns unstable after some mass has been lost. Alternatively, it is common practice to define an effective critical mass ratio $q \equiv M_{\rm a}/M_{\rm d}$, which is based on the full mass-transfer history. If the mass ratio at the onset of RLOF is larger than these critical values, mass transfer is stable. Due to it being an effective value, delayed dynamical instabilities, where the masses become more similar before the instability, are naturally taken into account. However, any set of critical mass ratios is only valid under a single set of assumptions regarding mass-transfer efficiency and angular momentum loss.

There are also difficulties of a numerical nature. For example, it is insufficient to assume that numerical breakdown of a 1D stellar-evolution code during mass transfer indicates physical instability.  Multiple physically motivated criteria have been proposed to identify unstable mass transfer. Indeed, there are multiple physical reasons why mass transfer might become unstable (see e.g. \citealt{Iva20}). In this section we describe different criteria that aim to identify unstable mass transfer. We later apply these different criteria to our simulations and, in Sect.~\ref{sec:results_qcrit}, compare the resulting critical mass ratios.

\subsection{Dynamical-timescale mass transfer}
\label{sec:crit_dyn}

A common way to determine whether RLOF is stable is by assuming that when the mass-transfer evolution accelerates to a dynamical timescale, such runaway mass transfer leads to the formation of a CE that engulfs both components of the binary. In this work, we compare the relevant timescales of both the donor star and the system as a whole to the binary period $P$, which we take as the characteristic timescale of the binary. We then consider a phase in binary evolution to be `dynamical' if the fractional amount by which either the mass of the donor star or the binary orbit changes over one orbital period exceeds a certain threshold. Specifically, we check whether
\begin{align}
\label{eq:dynev_crit}
    \max\left(\left| \dot{M}_{\rm d}/M_{\rm d} \right|,\, \left| \dot{a}/a\right|\right)\cdot P > A_{\rm dyn},
\end{align}
where $A_{\rm dyn}$ is a constant. 
We adopt $A_{\rm dyn} = 0.05$, which is in agreement with the start of the plunge-in phase as described in \cite{Iva20}. We then define the critical mass ratio for dynamical mass transfer, $q_{\rm dyn}$, as the lowest value of $M_{\rm a}/M_{\rm d}$ at the onset of RLOF for which the condition in Eq. \ref{eq:dynev_crit} is not met. We note that dynamical evolution does not always lead to runaway mass-transfer rates, and hence unstable mass transfer. In many cases, we still find self-regulating mass transfer phases for dynamically evolving binaries (see Sect. \ref{sec:results_mdot}).

We find that our results for $q_{\rm dyn}$ are sensitive to the precise value of $A_{\rm dyn}$, but not in equal measure over the full parameter space. The sensitivity to the value of $A_{\rm dyn}$ typically increases with decreasing initial donor mass or in more extended giants (see Sect. \ref{sec:results_qcrit} for more details). 

However, we consider that it would be misplaced to fixate too strongly on the precise choice of $A_{\rm dyn}$.  Once the binary is evolving on a dynamical timescale, the assumptions used to derive the Roche geometry themselves break down. The Roche potential is derived in the co-rotating reference frame, while neglecting the Euler force $\vec{f}_{\rm Euler} \propto \dot{\vec{\omega}}\times \vec{r}$, where $\omega$ is the angular velocity of rotation of the reference frame. Generally, this is an acceptable simplification. However, in the regime of dynamical evolution, the Euler force will compete with the other fictitious forces included in the potential, and so the Roche approximation is no longer valid. This is most easily recognised by considering the ratio of the Euler force to the centrifugal force:
\begin{align}
    \left|\vec{f}_{\rm Euler}/\vec{f}_{\rm centrifugal}\right| \sim \left|\frac{\dot{\omega}}{\omega^2}\right| = \frac{3}{4\pi} \left| \dot{a}/a\right| \cdot P~. 
\end{align}
With $A_{\rm dyn} = 0.05$ in Eq. \ref{eq:dynev_crit}, the Euler force contributes by at most about 1 per cent in systems that avoid dynamical evolution according to Eq. \ref{eq:dynev_crit}. Hence, once mass-transferring binaries are evolving dynamically, any results that assume the Roche geometry should be interpreted with care; this includes our own results.

\subsection{Overflow of an outer Lagrangian point}
\label{sec:crit_ol}

Alternatively, one could consider the overflowing of the equipotential surface passing through one of the outer Lagrangian points by the donor star as the onset of a dynamical instability. A similar criterion was adopted in \citet{Pac76}, the work commonly cited for the origin of the concept of CE evolution. Significant angular momentum loss through the outer Lagrangian points during mass loss (roughly the specific angular momentum of a circumbinary ring with radius $\sim 1.2 a$ for mass loss through $L_2$; see e.g. \citealp{Pri98}) is thought to dramatically shrink a binary's orbit and render the system dynamically unstable \citep[see e.g.][]{Nan14, Pav15}. However, we stress that the loss of mass through either of the outer Lagrangian points is not guaranteed to always result in unstable mass transfer \citep[see e.g.][for an observational example of this]{Bow10}. For example, under certain approximations, \citet{Lin17} find that any mass loss through the $L_2$ point occurs at a lower rate than conventional mass loss through the $L_1$ point and that its effects on the overall stability of a binary could be only marginal. Nevertheless, donor stars that extend past their OL, which passes through the $L_3$ point if $M_{\rm a} \leq M_{\rm d}$ and otherwise through the $L_2$ point, significantly violate the assumption of spherical symmetry made in 1D evolution codes, which makes reliably simulating this phase challenging. 

We do not model mass loss through outer lobe overflow (OLOF), and hence ignore the effects thereof on the evolution of our binaries (but see \citealt{Marchant+2021} and Bobrick et al. in prep.). However, we keep track of whether or not the donor star experiences OLOF by comparing its radius to the equivalent-volume radius of the OL, $R_{\rm OL}$. To that end, we numerically calculated $R_{\rm OL}$ using \texttt{Fortran} routines for $-4 \leq \log q \leq 4$, including a damped Newton-Raphson method to find the location of the outer Lagrangian point, and double Simpson method to calculate the volume of the potential surface passing through this point. The results we obtained are converged to better than one part in $10^8$. A Monte Carlo method written in \texttt{Python} was used to extend our calculations of $R_{\rm OL}$ beyond our \texttt{Fortran} routines to $-7 \leq \log q \leq 7$. We fit the ratio of the equivalent-volume radius of the OL to that of the L$_{\rm 1}$ lobe with the following function:
\begin{align}
    \frac{R_{OL}}{R_{L_1}} = 1 + \frac{0.441q^{-0.325}}{1 + 0.412q^{-0.8}}.
\end{align}
This fit is accurate to better than $0.2\%$ over the full range of $q$ values. Analogous to the previous section, we define a critical OLOF mass ratio $q_{\rm OLOF}$ as the lowest value of $\left.M_{\rm a}/M_{\rm d}\right|_{\rm RLOF}$ for which the donor star does not experience OLOF during the entire mass transfer. Typically, we find that the overflow of outer Lagrangian points occurs in systems where the evolution becomes dynamical, meaning that $q_{\rm dyn}$ and $q_{\rm OLOF}$ have similar values (see Sect.~\ref{sec:results_qcrit}).

\subsection{Thermal readjustment of surface layers}
\label{sec:crit_th}

\begin{figure}
\centering
\includegraphics[width=\hsize]{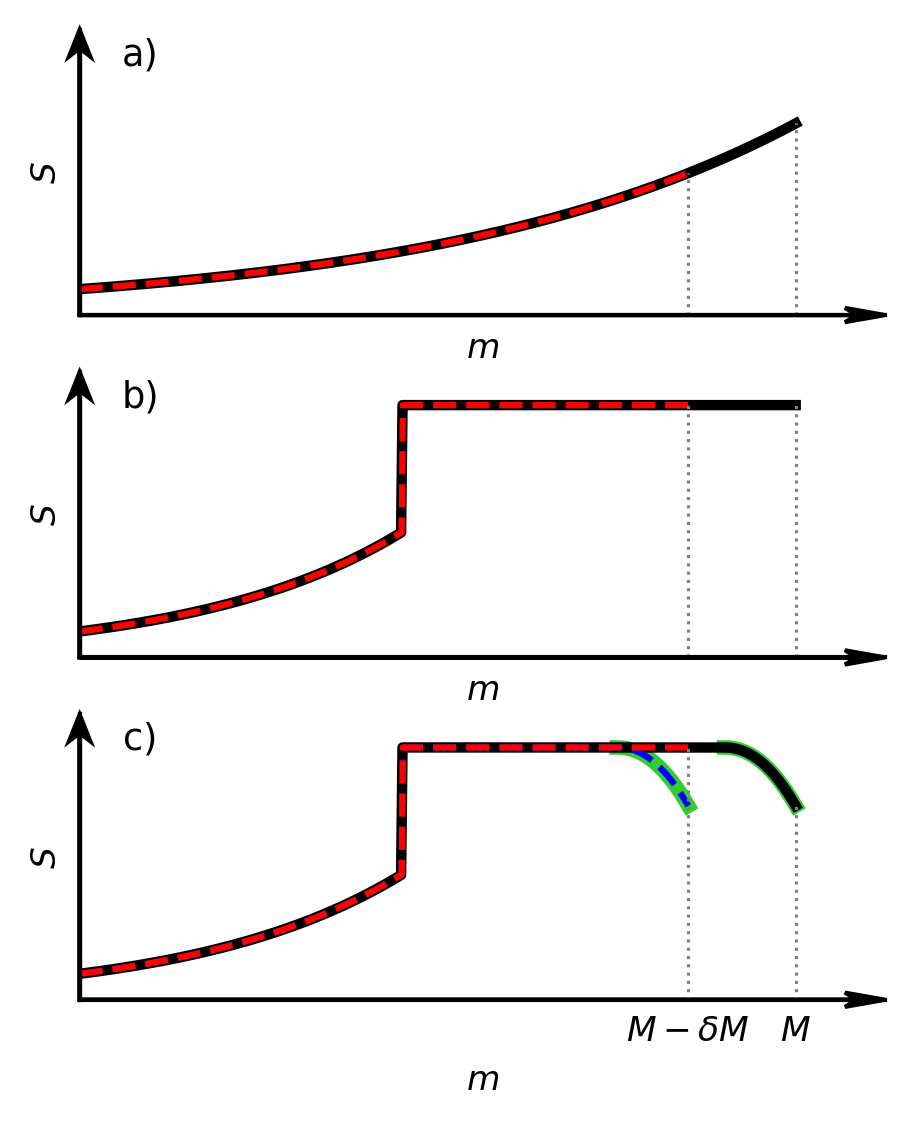}
  \caption{Schematic illustration of the entropy ($S$) profile versus mass coordinate $m$ in various types of stars. Panel a) depicts a representation of a star with a radiative envelope, whilst panel b) shows a model for a giant with a convective, isentropic envelope. A more realistic representation of the envelope of a convective giant is shown in panel c), which includes the superadiabatic sub-surface layers (highlighted in green). In each panel, the black line represents the entropy profile of the undisturbed star. Dashed red lines represent the entropy profile after sudden adiabatic (see text) loss of $\Delta M$ of mass. The dashed blue line in panel c) shows the entropy profile if the mass is lost at a rate below the critical mass-loss rate (Eq. \ref{eq:mdot_crit_max}).}
     \label{fig:entropy_mass_transfer_response}
\end{figure}

\begin{figure}
\centering
\includegraphics[width=\hsize]{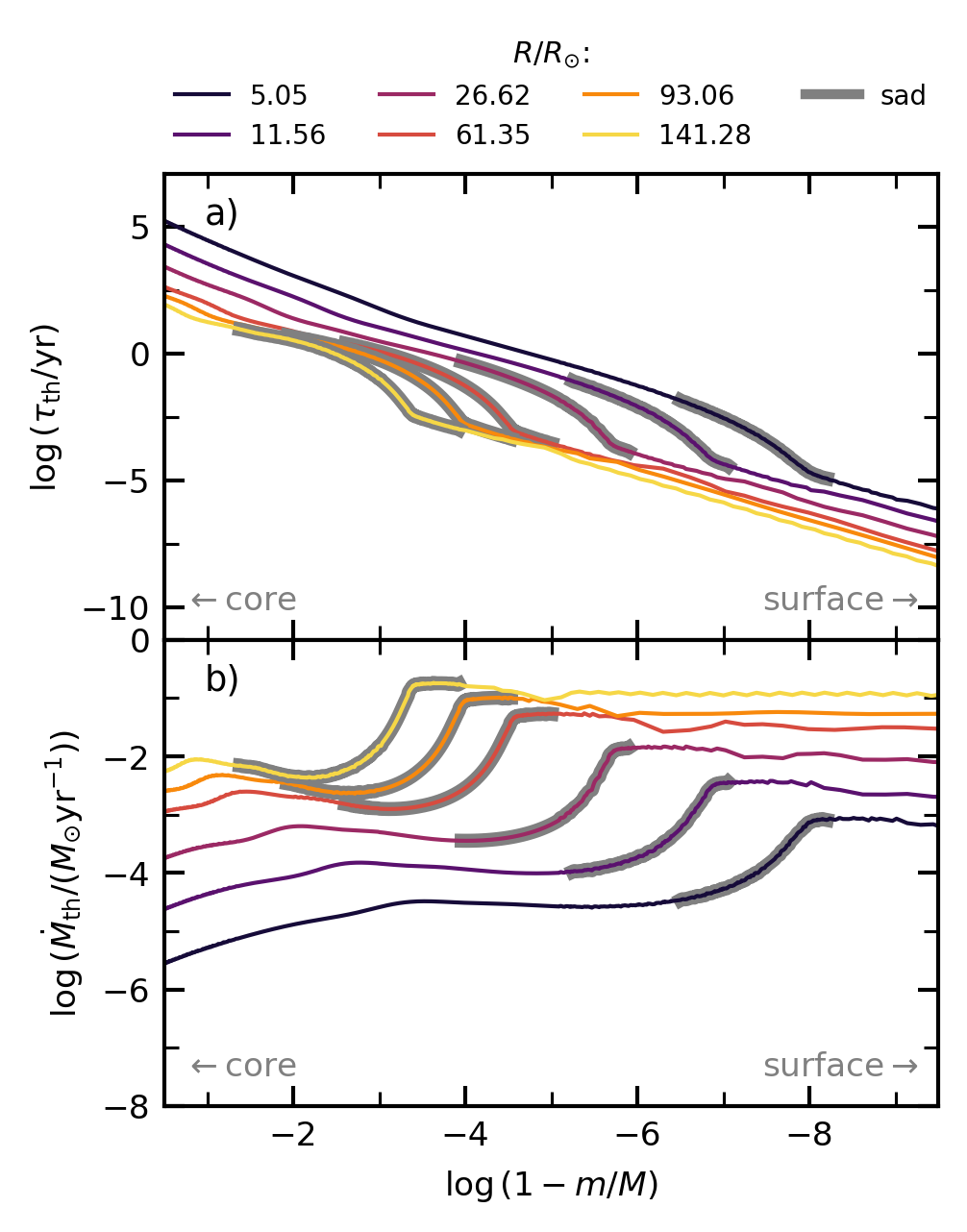}
  \caption{Thermal properties of a $1.5\, M_\odot$ donor star at various stages in its post-MS evolution. The colours of the lines correspond to the moments the star reaches certain radii (as shown in Fig. \ref{fig:model_selection}), whose values are shown in the legend. Regions with grey shading are significantly superadiabatic (`sad'; here defined as regions where $\nabla - \nabla_{\rm ad} \geq 0.1 \nabla$; with $\nabla = \frac{\partial \log T}{\partial \log P}$). The top panel shows the run of the local thermal timescale (see Eq. \ref{eq:tau_th_loc}) with the logarithm of the fractional exterior mass, such that the stellar centre is located on the left end of the x axis, and the surface is located to the right. The bottom panel shows the run of the associated thermal mass-loss rate (see Eq.~\ref{eq:mdot_crit}) on the same horizontal axis. }
     \label{fig:sstar_mdot_crit}
\end{figure}

\begin{figure}
\centering
\includegraphics[width=\hsize]{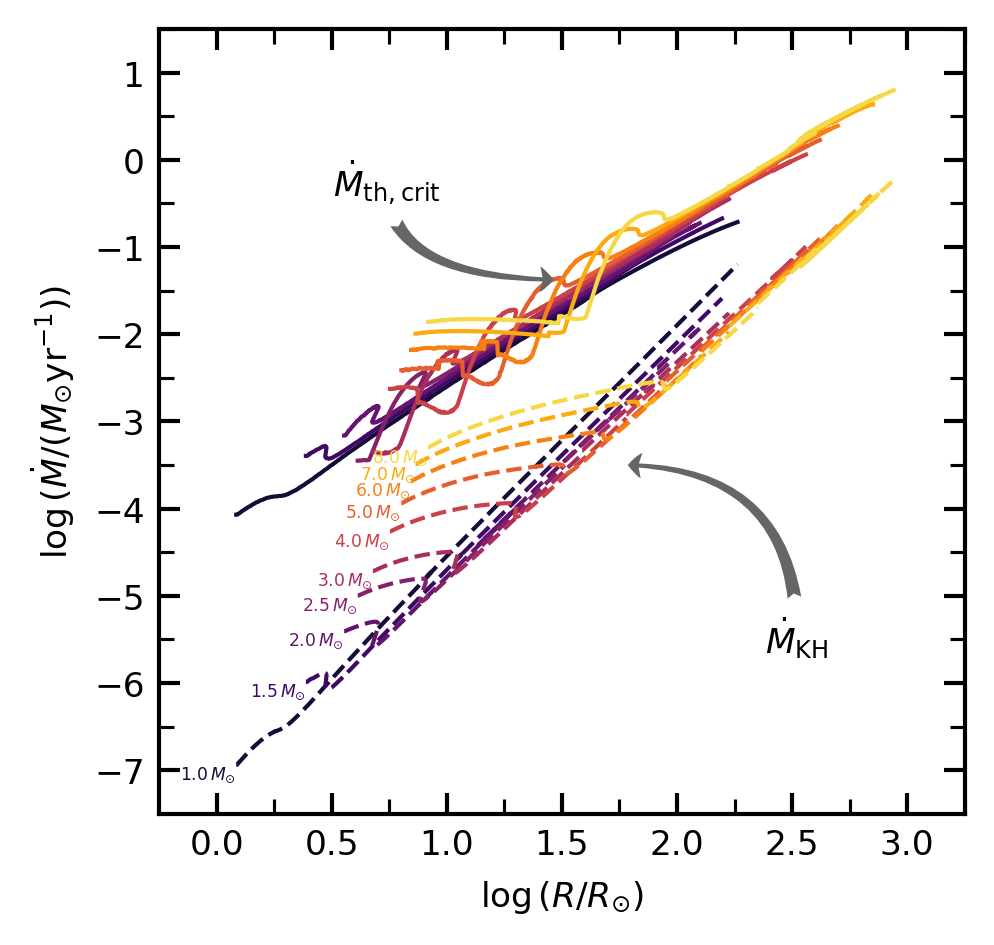}
  \caption{Critical mass-loss rate as a function of stellar radius from thermal considerations (see Eq. \ref{eq:mdot_crit_max}). Solid lines show the critical mass-loss rate based on local thermal properties of the sub-surface layers of the donor star (see Eq. \ref{eq:mdot_crit} and Fig. \ref{fig:sstar_mdot_crit}), whilst dashed lines show the critical mass-loss rate based on global thermal properties (see Eq. \ref{eq:mdot_KH}). The linear portions of the curves correspond to the giant phases. The colours of the lines correspond to different initial donor masses, as indicated to the lower left of each dashed line.}
     \label{fig:sstar_mdot_crit_global}
\end{figure}

Directly following a sudden loss of mass, the donor star loses both its hydrostatic and thermal equilibria. In the simplified picture, the initial response of the donor star occurs on the dynamical timescale, when it attempts to restore its hydrostatic equilibrium. After that, mass transfer is driven by the slower thermal readjustment of the donor, as the donor star attempts to recover the thermal equilibrium radius appropriate for its new mass on the global (KH) thermal timescale:
\begin{align}
\label{eq:tau_th_kh}
\tau_{\rm KH} = \frac{1}{2} G\frac{M_{\rm d}^2}{RL} \sim  1.5 \cdot 10^7 \frac{M_{\rm d}^2}{M_\odot^2} \frac{R_\odot}{R} \frac{L_\odot}{L}\; \text{yr}.
\end{align}
A very simple estimate for the rate of mass transfer occurring on a donor star's global thermal timescale is then given by
\begin{align}
\label{eq:mdot_KH}
    \dot{M}_{\rm KH} \equiv -\frac{M_{\rm d}}{\tau_{KH}} \sim -6.7\cdot 10^{-7}  \frac{M_\odot}{M_{\rm d}} \frac{R}{R_\odot} \frac{L}{L_\odot}\; M_\odot \text{yr}^{-1}.
\end{align}

More specifically, in the simplified picture it is assumed that the donor star initially responds to the loss of mass adiabatically. With this adiabatic assumption, the new sub-surface layers in a radiative star (shown as the red dashed curve in Fig.~\ref{fig:entropy_mass_transfer_response}a) will have lower entropy compared to the sub-surface layers before the loss of mass. Hence, after adiabatic decompression, these layers will occupy a smaller volume than the layers that have been removed, which stabilises the mass transfer and leads to relatively low critical mass ratios. In the envelope of a convective giant, approximated as isentropic, the new sub-surface layers (shown as the red dashed curve in Fig.~\ref{fig:entropy_mass_transfer_response}b) will have similar entropy as before the loss of mass. This results in the sub-surface layers expanding along with the star as a whole in their response to the loss of mass \citep[we note that even in the context of polytropic models, the conclusion that an isentropic envelope expands upon mass loss depends on the fractional core mass;][]{Hje87}. This behaviour of these convective giants is reflected in their generally comparatively large critical mass ratios. Typically, the location of the boundary between stable and unstable mass transfer used in prescriptions used in rapid BPS studies \citep[see e.g.][]{Too12,Cla14,compass22} is based on model calculations \citep[such as][]{Hje87,Ge10} assuming (part of) these simplifications.

However, even on timescales much shorter than the global thermal timescale, stars do not necessarily react to the loss of mass fully adiabatically, since surface layers respond more rapidly than the star as a whole \citep[see e.g.][]{Osa70, Pos02, Kip12}. As emphasised by \citet{Woo11} and \citet{Pas12}, the thermal readjustment of the surface layers of the donor star increases mass-transfer stability as compared to adiabatic models (see more below). The different thermal timescales on which the different sub-surface layers of a donor star will react depend on the local properties of these layers, for example the efficiency of convection, the degree of ionisation, and the specific energy content \citep[see also][]{Woo11,Pas12,Pav15}. Approximating the local thermal timescale at mass coordinate $m$ as the time it would take a star to radiate away all the thermal energy content of the layers above $m$ at the rate of its surface luminosity, the local thermal timescale of a given layer can be calculated as \citep[e.g.][]{Kip12}
\begin{align}
\label{eq:tau_th_loc}
\tau_{\rm th}(m) = \frac{1}{L}\int_m^M c_P(\tilde{m})T(\tilde{m}) \text{d}\tilde{m}.
\end{align}
We note that the above equation implicitly assumes that one can divide the stellar interior into an outer and an inner zone at any mass coordinate, and that the outer zone can thermally readjust independently from the inner zone. This assumption might be inappropriate, especially in a convective region, where the timescale of convective overturn could be shorter than the local thermal timescale, which could stabilize thermal perturbations on a shorter timescale.

We present runs of the local thermal timescale with the fractional exterior mass for different moments in the evolution of a $1.5~M_\odot$ donor star in the upper panel of Fig \ref{fig:sstar_mdot_crit}. Similar to the global (KH) thermal timescale, at any mass coordinate relative to the surface the local thermal timescale tends to decrease as the star expands and becomes more luminous. Additionally, the local thermal timescale decreases monotonically with the mass coordinate $m$. However, there is a part of the sub-surface layers where the stellar interior is unstable to convection, but energy transport via convection is inefficient. This results in temperature gradients that are significantly larger than the adiabatic temperature gradient, and these regions are hence known as superadiabatic regions (here defined as regions where $\nabla - \nabla_{\rm ad} \geq 0.1 \nabla$; with $\nabla = \frac{\partial \log T}{\partial \log P}$). Within these superadiabatic sub-surface layers (highlighted by black shading in Fig. \ref{fig:sstar_mdot_crit}), the steep temperature gradients result in a relatively sharp decrease in the local thermal timescales with $m$ in the upper parts of these regions, compared to above and below the superadiabatic layers. Donor stars do not respond effectively adiabatically to mass loss until a certain critical mass-transfer rate is reached, determined largely by these relatively steep slopes of $\tau_{\rm th}(m)$ in the upper part of the superadiabatic regions. 

We can make this more explicit by examining the thermal mass-loss rate $\dot{M}_{\rm th}(m)$ as a function of mass coordinate. This is the minimum mass-loss rate required to strip away the outer layers of a star down to this mass coordinate $m$ faster than these layers can thermally readjust. This mass-loss rate can be estimated in the following way:
\begin{align}
\label{eq:mdot_crit}
    \dot{M}_{\rm th}(m) = -\frac{M-m}{\tau_{\rm th}(m)},
\end{align}
which is plotted in the bottom panel of Fig \ref{fig:sstar_mdot_crit}. It can be seen that $\dot{M}_{\rm th}(m)$ generally reaches its maximum value 
\begin{align}
\label{eq:mdot_crit_max}
    \dot{M}_{\rm th,crit} = \max \dot{M}_{\rm th} (m)
\end{align}
at the top of the superadiabatic layer. As long as the mass-loss rate does not exceed $\dot{M}_{\rm th, crit}$, there is always a part of the outer layers that can thermally readjust on a timescale shorter than the mass loss timescale. Conversely, if the mass-loss rate should exceed this critical value, no layer would be able to thermally adjust on short enough timescales, and the mass transfer can be expected to take place fully adiabatically. 

In the outer sub-surface layers, the local thermal timescales are shorter than the donor star's global and local dynamical timescales. Hence, for mass-transfer rates up to $\dot{M}_{\rm th, crit}$, the initial response of a convective giant is dictated by the readjustment of the local thermal structure. In this scenario, the new sub-surface layers (shown as the blue dashed curve in Fig.~\ref{fig:entropy_mass_transfer_response}c) have a lower entropy than before the mass loss. Thus, similar to what happens in a radiative star, the sub-surface layers occupy a smaller volume after decompression. This stabilises mass transfer over a significantly wider range of initial mass ratios than when the possibility of thermal relaxation is ignored. However, if the mass loss becomes too rapid ($\dot{M}_{\rm d} > \dot{M}_{\rm th, crit}$), the donor star is not able to readjust to the loss of mass on its local thermal timescale, and its response becomes adiabatic. In that scenario, the new sub-surface layers (shown as the red dashed curve in Fig. \ref{fig:entropy_mass_transfer_response}c) will have a higher entropy, compared to the ones before the mass loss. The sub-surface layers thus expand relative to the star as a whole, raising the mass-transfer rate even more and driving the mass transfer to instability. 

As such, the transition to fully adiabatic response of the donor star is an indication for the onset of unstable mass transfer. We thus keep track of $\dot{M}_{\rm th, crit}$ during the evolution of our binary systems, and whether or not this value is exceeded by $\dot{M}_d$. We define a last, 'quasi-adiabatic' critical mass ratio $q_{\rm qad}$ as the lowest value of $\left.M_{\rm a}/M_{\rm d}\right|_{\rm RLOF}$ for which we can simulate the complete mass transfer phase with MESA and the mass-transfer rate does not exceed $\dot{M}_{\rm th, crit}$. 

In Fig.~\ref{fig:sstar_mdot_crit_global} we show how $\dot{M}_{\rm th, crit}$ changes during the evolution of donor stars with different ZAMS masses. Like the global KH mass-transfer rate, $\dot{M}_{\rm th, crit}$ typically increases with donor radius. This is a direct consequence of the local thermal timescale at any depth decreasing with the stellar radius. At a given donor radius, $\dot{M}_{\rm th, crit}$ typically increases with donor mass, whilst $\dot{M}_{\rm KH}$ typically decreases with donor mass. For the least evolved stars, $\dot{M}_{\rm KH}$ is up to about three orders of magnitude lower than $\dot{M}_{\rm th, crit}$. However, since $\dot{M}_{\rm KH}$ is a steeper function of $R_{\rm d}$ than is $\dot{M}_{\rm th, crit}$, the two rates grow more similar with increasing donor radius.

As briefly mentioned earlier, we do not find that dynamical or OLOF evolution always leads to runaway, unstable mass transfer. In fact, mass transfer is self-regulating in most of the systems with dynamical or OLOF evolution within the context of our models, provided they did not first encounter the quasi-adiabatic criterion. By contrast, we find that our quasi-adiabatic criterion, based on local thermal properties of the donor star, does consistently pick out the systems with runaway mass transfer. Hence, in this work, we base our main results on the quasi-adiabatic criterion described in Sect. \ref{sec:crit_th}. However, we also show the effects of choosing different criteria, as described in Sects. \ref{sec:crit_dyn} and \ref{sec:crit_ol}.

\section{Results}
\label{sec:results}

In this section we present the results from our binary simulations. In Sect. \ref{sec:results_mdot}, we focus on the evolution of the mass-transfer rates in our binaries, whilst we present the critical mass ratios for stable mass transfer in Sect. \ref{sec:results_qcrit}. Our main results are tabulated in Appendix B.

\subsection{Mass-transfer rates and evolution}
\label{sec:results_mdot}

\begin{figure*}
\centering
\includegraphics[width=\hsize]{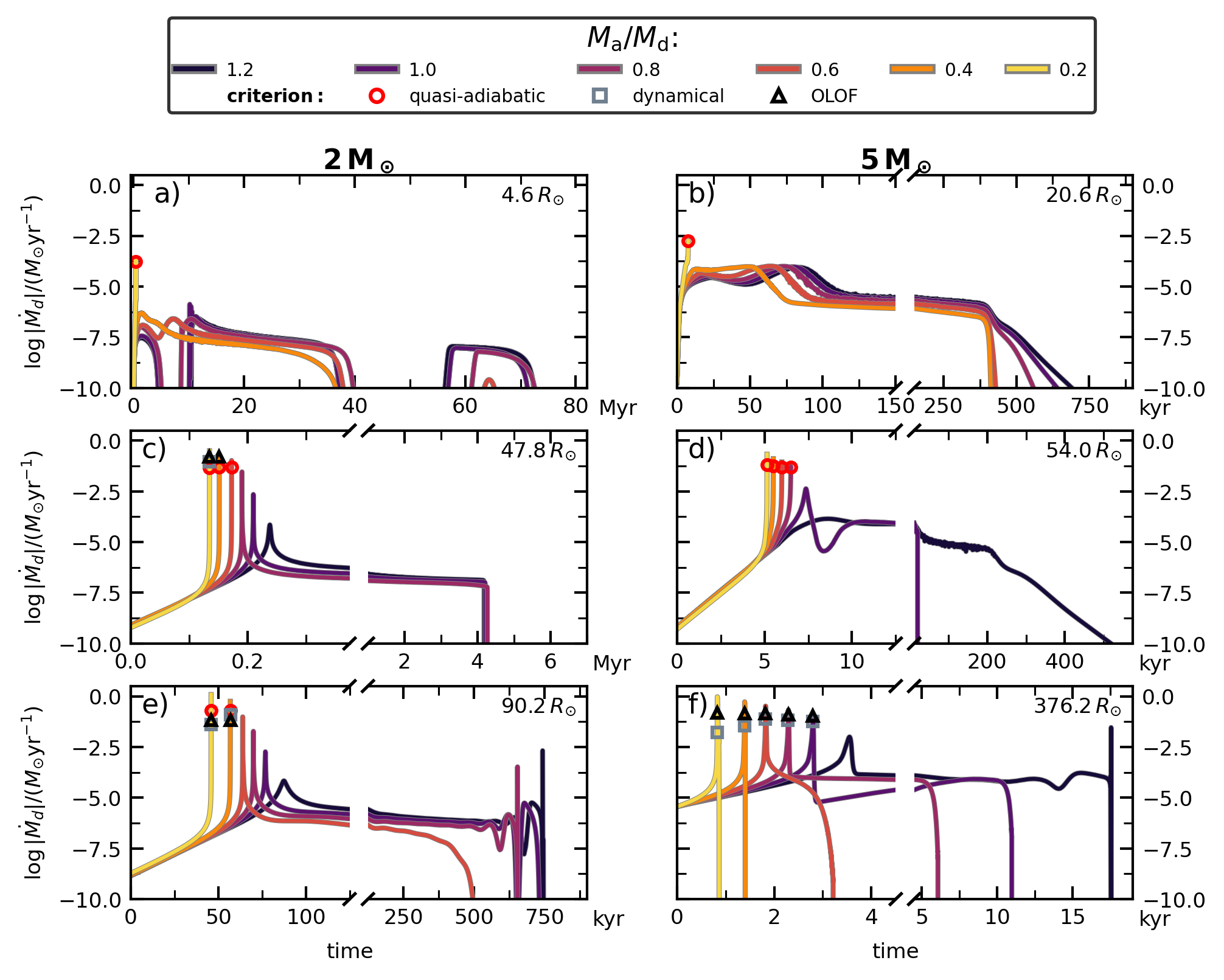}
   \caption{Time evolution of mass-transfer rates in systems with $M_{\rm d}=2\,M_\odot$ (left column) and $M_{\rm d}=5\,M_\odot$ (right column). Each row corresponds to RLOF starting in a different phase of the donor stars' lives (at the radius indicated in the top right of each panel). More specifically, panels a) and b) show HG donors, panels c) and d) show donors on their RGB, and panels e) and f) show donors on their AGB. The colours of the lines correspond to different mass ratios at the onset of RLOF, as indicated in the legend. We have indicated where, if appropriate, our quasi-adiabatic-, dynamical-, and OLOF-based criteria are met with open circles, squares, and triangles, respectively. The zero point of the time axis is defined as the first moment when the mass-transfer rate exceeds the rate at which wind mass loss occurs. }
     \label{fig:mdot-hists}
\end{figure*}

\begin{figure}
\centering
\includegraphics[width=\hsize]{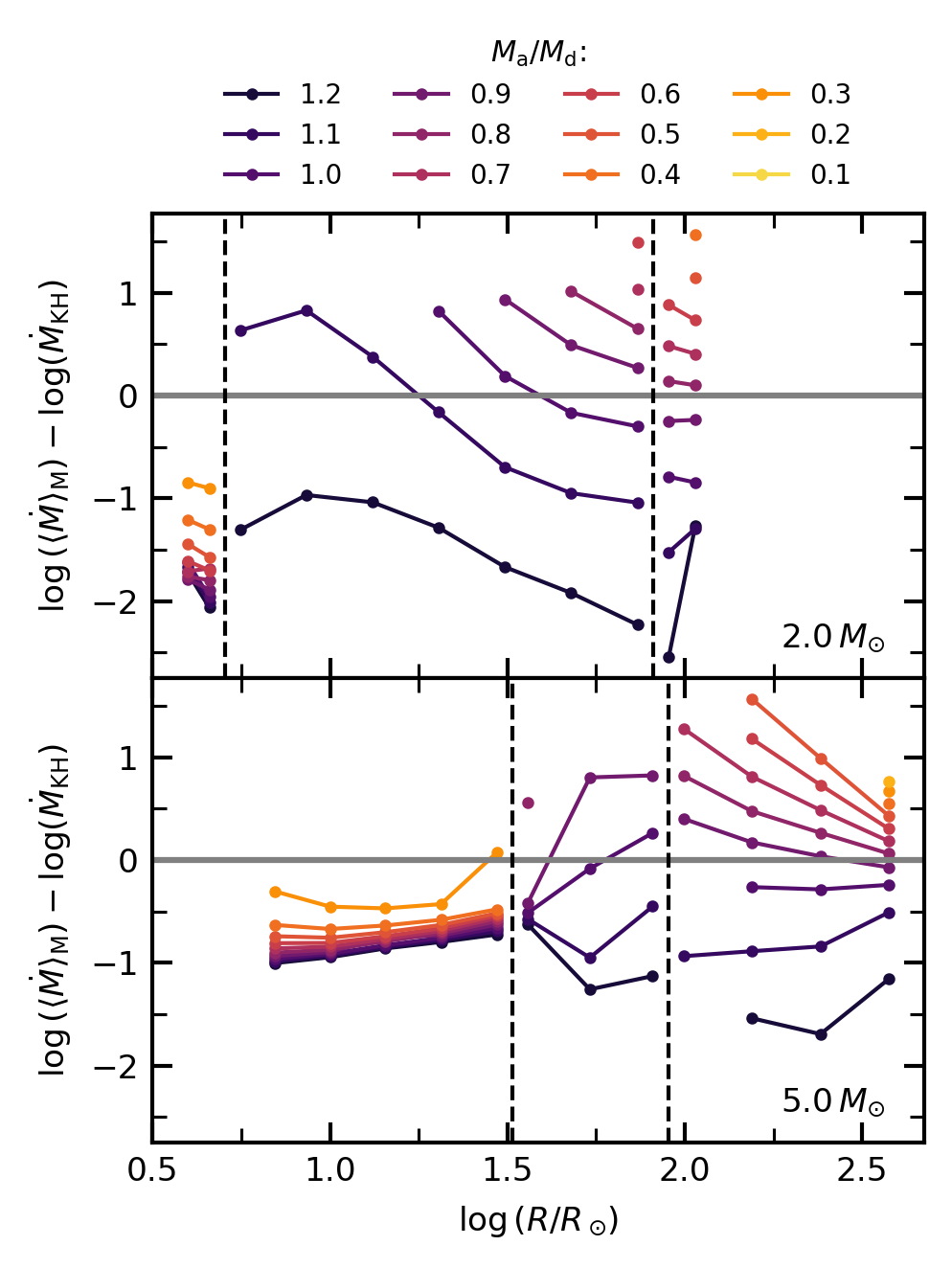}
  \caption{Mass-transfer rates from our calculations compared to the global KH rate (as given by Eq. \ref{eq:mdot_KH}). The top panel shows results for a $2~M_\odot$ donor star, whilst the bottom panel shows results for a $5~M_\odot$ donor star. In each panel, the coloured lines show the difference between the mass-transfer rate predicted by Eq.\,\ref{eq:mdot_KH} (which assumes the rate is governed by the global KH timescale at the onset of mass transfer) and the mass-averaged mass-loss rate $\langle\dot{M}_{\rm d}\rangle_{\rm M}$ (see Eq. \ref{eq:mass_ave_mdot}) from our calculations. Vertical dashed black lines indicate the base and the tip of the RGB. }
     \label{fig:mdot-mdotKH_compare}
\end{figure}

For most of the systems that we simulated, the overall evolution of the mass-transfer rate can be divided into two distinct parts (see Fig. \ref{fig:mdot-hists} for a few examples). The first is a relatively brief `thermal' phase, when the donor star attempts to recover thermal equilibrium and where the mass-transfer rate peaks. This is followed by a significantly longer phase, driven by the natural expansion of the donor star on its current evolution timescale, where the mass transfer occurs at lower rates. The mass-transfer rates depend on the parameters of the systems at the onset of RLOF, in that both more evolved and more massive donor stars will typically lose mass at higher rates. This trend is a result of the global thermal timescale decreasing with increasing luminosity and radius (Eq. \ref{eq:tau_th_kh}). 

In the following, we discuss our results for Hertzsprung gap (HG) donors separately from those for giant donor stars. In this work, we use the term HG to refer to all stars evolving from the terminal-age main sequence to the base of 
the RGB. These stars typically have radiative envelopes and evolve on their thermal  timescales. Unless explicitly stated otherwise, we also include in that term low-mass subgiants that expand on a much slower, near-nuclear timescale.

\subsubsection{Hertzsprung gap donors}
\label{sec:hg_donors}
Due to the structure of a radiative envelope and a relatively long global thermal timescale, the initial response of a HG donor to mass loss is to shrink. Because such a donor star needs to expand to comparatively large radii to regain its thermal equilibrium, the mass transfer is driven on the thermal timescale of the donor. 

Our interests lie predominantly in the first, rapid part of the mass transfer phase, since any potential instabilities will occur there. We use the mass-averaged mass-loss rate
\begin{align}
    \label{eq:mass_ave_mdot}
    \langle\dot{M}_{\rm d}\rangle_{\rm M} = \frac{1}{M_{\rm d,i} - M_{\rm d,f}}\int_{M_{\rm d,i}}^{M_{\rm d,f}} \dot{M}_{\rm d}(M_{\rm d}) \text{d}M_{\rm d},
\end{align}
 as a representative mass-transfer rate. The subscripts `f' and `i' correspond to quantities evaluated at the final and initial moments in the mass-transfer evolution (here defined as the moments where $ \dot{M}_{\rm d} \geq 10^{-15} M_\odot$yr$^{-1}$). The rapid thermal part is short compared to the subsequent mass transfer on the evolutionary timescale of the donor star, and hence the time-averaged mass-transfer rate would be dominated by the latter. The mass-averaged mass-loss rate, as defined above, does not suffer from this issue, as most mass is lost during this rapid phase.  
 
 Figure \ref{fig:mdot-mdotKH_compare} compares $\langle\dot{M}_{\rm d}\rangle_{\rm M}$ to the KH rate (as given by Eq. \ref{eq:mdot_KH}) for $2~M_\odot$ and $5~M_\odot$ donor stars. It is clear that donors that fill their Roche lobes as HG stars typically lose mass at rates that are significantly below $\dot{M}_{\rm KH}$, by up to a factor of 10. This weakly depends on the mass ratio, and is almost independent of the orbital period for the more massive donors that expand significantly during the HG. Similarly, the peak mass-transfer rates, whilst typically closer, are often below $\dot{M}_{\rm KH}$, though higher than $\langle\dot{M}_{\rm d}\rangle_{\rm M}$ by a factor of a few. 
 
 For HG donors, the evolution of the mass transfer can be fairly complicated. Often, the rapid phase of mass transfer consists of two peaks instead of one. At the start of mass transfer, the initially radiative donor star shrinks slightly as the radius follows the Roche radius. However, shortly after the first peak in the mass-transfer rate, the convective zone in the outer layers of the donor star extends increasingly far down into the stellar interior. The mass-transfer rate climbs to a second peak because these convective outer layers can supply the required material faster than a radiative envelope. Typically, the second peak coincides with the moment the convective envelope has grown to its largest extent \citep[see also e.g. ][]{Har70, Lin87}. 

For many $M_{\rm d} = 2-3~M_\odot$ donors with $q \ga0.7$  that start transferring mass before the base of the giant branch, the aforementioned shrinkage is significant enough to temporarily suspend mass transfer altogether (see Fig. \ref{fig:mdot-hists}a). Typically, such interruptions last of the order of a few megayears, equivalent to a few per cent of the total duration of the mass transfer, or a few times the global thermal timescale of the donor stars at that point.

In several $M_{\rm d} \leq 2~M_\odot$ donors, the convective part of the envelope extends down deeply enough to leave a discontinuity in the abundance profile at the deepest point of the convective zone. When a star's H-burning shell reaches this discontinuity, the star's luminosity and radius decrease slightly. This process also occurs in single stars \citep[e.g.][]{Tho67, Ibe68, Chr15}, where it is well-known to result in the so-called `luminosity bump' in star clusters \citep[e.g.][]{Kin85, Sal02}. For the aforementioned low-mass donors, the associated shrinkage can be significant enough to suspend the mass transfer until the donor re-expands and once more fills its Roche lobe \citep[as already shown by][]{Kip67}, of the order of tens of megayears later.

Thus, several of the $2~M_\odot$ HG donors experience two interruptions in their mass transfer, as shown in Fig. \ref{fig:mdot-hists}a \citep[see also e.g.][]{Han00}. By contrast, for the more massive stars ($M \geq 4~M_\odot$), mass transfer is not suspended at any moment during the mass transfer in our models. These stars hardly become convective during mass loss (the mass of the convective part of the outer layers is typically only $\sim 0.1~M_\odot$), and therefore they do not undergo significant shrinkage induced by either structural re-arrangement or by dredge-up/mixing effects.

In Fig. \ref{fig:mdot-hists} we have indicated with different symbols when during the evolution of our binaries our three instability criteria are met, if applicable. For HG donor stars, none of the systems we simulated experienced dynamical or OLOF evolution before $\dot{M}_{\rm th,crit}$ was reached. By contrast, we do find that our quasi-adiabatic criterion (see Sect. \ref{sec:crit_th}) is consistently met for systems where the mass-transfer rate runs away. Furthermore, we find that mass transfer does not become unstable immediately after the onset of RLOF, but that the binaries instead experience a delayed instability \citep[see e.g.][]{Hje87,Han06,Pav15, Ge20a}. The amount of time it takes for mass transfer to become unstable is generally a function of the mass ratio, with the typical delay time being shorter in systems with more extreme mass ratios.

We sometimes find systems with anomalous mass transfer behaviour along the boundary between stable and unstable mass transfer. During the rapid thermal part of the mass transfer, $\dot{M}_{\rm d}$ oscillates by up to three orders of magnitude, with an oscillation period similar to the typical critical thermal timescale during that phase. During several of the peaks in the mass-transfer rate, the rate at which the donor star loses mass exceeds the critical thermal rate (see Sect. \ref{sec:crit_th}). In these moments, the outermost layers are stripped away quasi-adiabatically (as the surface layers readjust on a longer timescale than the timescale of mass loss), and the mainly radiative regions below the convective outer layers briefly approach the surface. The donors recover as the mass-transfer rate decreases and the convective part of the outer layers regrows, restarting the cycle anew. After a few of these cycles, the mass-transfer rate stabilises and follows the typical evolution described in this section.

\subsubsection{Giant donors}
\label{sec:giant_donors}

Donor stars that fill their Roche lobes as giants typically have convective outer layers. Convective envelopes are effectively isentropic (excepting the non-negligible superadiabatic region) and therefore tend to expand relative to their Roche lobe upon mass loss, resulting in higher, but shorter-lasting peaks in the mass-loss rates. More evolved giant donors have shorter evolutionary timescales and therefore higher mass-transfer rates, as can be seen in Fig.~ \ref{fig:mdot-hists}. For the majority of giant donors in our grid, we find that general properties of the mass-transfer evolution (such as duration and typical or peak mass-transfer rates) are poorly captured by Eq.~\ref{eq:mdot_KH}. This is because this type of formula does not depend on properties of the binary systems, such as the mass ratio or the amount by which the donor overflows its Roche lobe, but only on global properties of the donor star \citep[see also e.g.][who find a similar mismatch]{Lan00}. 

Unlike for HG donor stars, $\langle\dot{M}_{\rm d}\rangle_{\rm M}$ for giant donors is strongly dependent on the initial mass ratio and period of a binary (see Fig.~\ref{fig:mdot-mdotKH_compare}). This is especially true for the low-mass giants ($M_{\rm d} \leq~2.5~M_\odot$), where $\langle\dot{M}_{\rm d}\rangle_{\rm M}$ can range from over two orders of magnitude below to almost two orders of magnitude above the global thermal rate of the donors. For the more massive giants, the dependence on the initial mass ratio is less extreme. The peak mass-transfer rate follows a very similar trend with mass ratio and period, being about an order of magnitude higher than $\langle\dot{M}_{\rm d}\rangle_{\rm M}$ for unevolved giants, with the difference decreasing for more evolved giant donors. Over the full mass range, the mass-averaged mass-transfer rate tends to agree better with the global thermal rate, and depends less strongly on mass ratio, with increasing donor radius. As we describe in more detail below, even in the systems for which $\langle\dot{M}_{\rm d}\rangle_{\rm M} \gg \dot{M}_{\rm KH}$, mass transfer is still self-regulating as long as the critical mass-transfer rate (see Sect. \ref{sec:crit_th}) is not exceeded.

As for the less evolved donors discussed in Sect. \ref{sec:hg_donors}, we have marked in Fig. \ref{fig:mdot-hists} when the binaries with giant donors meet our instability criteria, if applicable. As Fig. \ref{fig:mdot-hists}f clearly illustrates, neither dynamical-timescale nor OLOF evolution necessarily lead to unstable, runaway mass transfer. As we noted for the HG donors, the mass transfer does typically turn unstable when the critical thermal rate (as given in Eq. \ref{eq:mdot_crit_max}) is exceeded. As such, the dynamical and OLOF criteria should be interpreted more as guidelines for when our 1D simulations start becoming unrealistic, rather than actual physical stability criteria. This is discussed further in Sect. \ref{sec:results_qcrit}. 

Typically, stable mass transfer for giant donors continues until nearly the entire envelope has been stripped. However, for relatively unevolved RGB stars where helium is ignited under non-degenerate conditions (in our grid this corresponds to $M_{\rm d} \geq 2.5~M_\odot$; $\log T_{\rm eff, d} \gtrsim 3.6$) with mass ratios that are not too extreme (${M_{\rm a}}/{M_{\rm d}} \gtrsim 0.8$), the donor star can start burning helium in its core during mass transfer when there is still a significant amount of mass left in the envelope. As a result, after shrinking within their Roche lobes, which suspends the mass transfer, they evolve through the HRD in a bluewards loop within the HG region, without becoming hotter than the main-sequence band. The donor stars can re-expand enough to refill their Roche lobes as partially stripped AGB stars. The typical amount of mass left in the envelope is around 30--40\% of the total donor mass at that point. 

Lastly, we find that the most evolved AGB donors in our grid can experience TPs whilst transferring mass. As a result of recurring He shell flashes, these donor stars experience cycles of expansion and contraction. Consequently, also the mass-transfer rate increases and decreases significantly during these cycles, as visible in Fig. \ref{fig:mdot-hists} panels e) and f). For many donor stars, the shrinkage is significant enough to completely suspend mass transfer temporarily. Additionally, for several systems, we find that the expansion is large enough to drive mass transfer to higher rates than during the rapid peak after the onset of RLOF. 

\subsection{Critical mass ratios}
\label{sec:results_qcrit}
\begin{figure}
\centering
\includegraphics[width=\hsize]{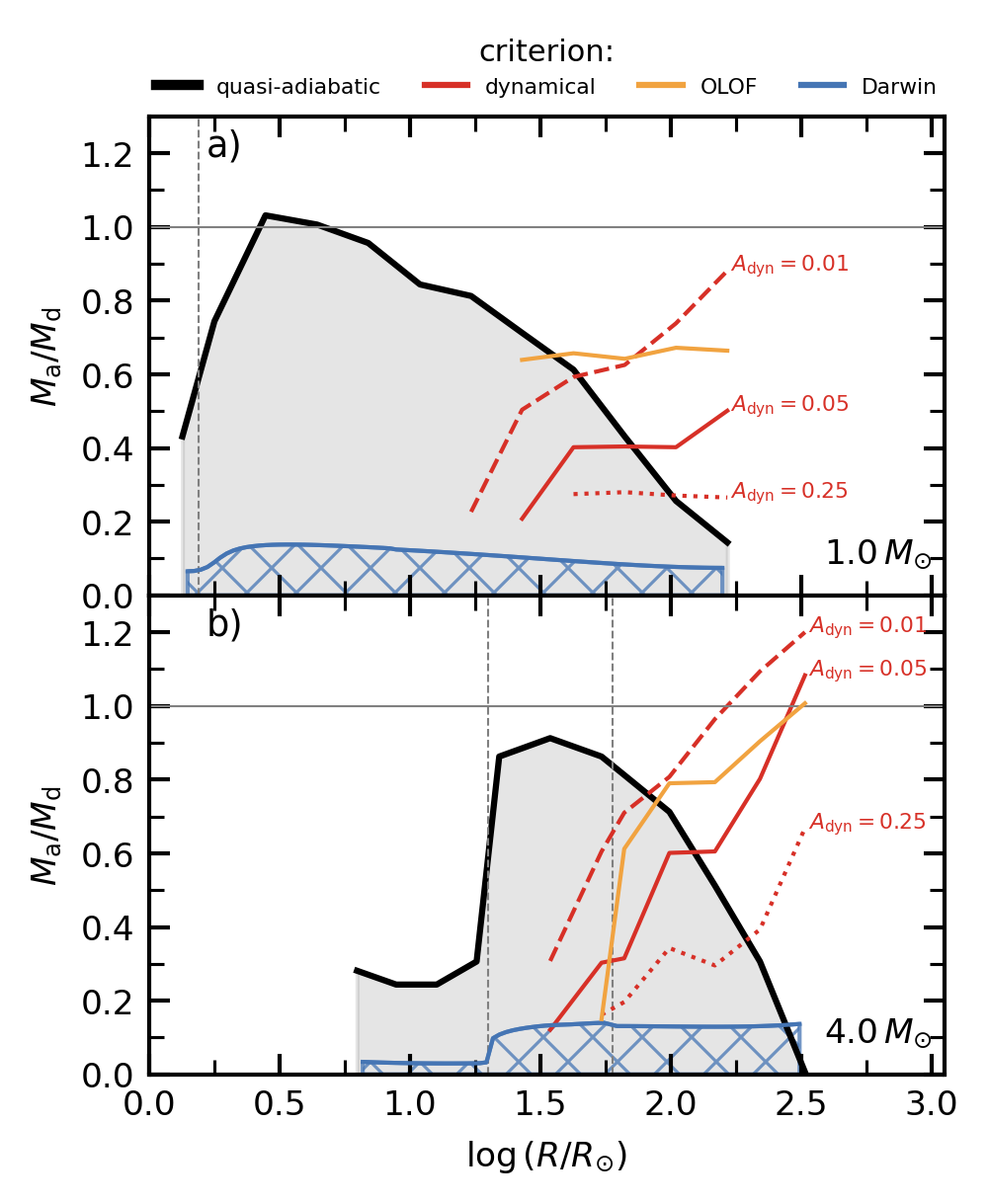}
  \caption{Critical mass ratios (above which mass transfer is stable) for $1~M_\odot$ and $4~M_\odot$ donor stars as a function of donor radius, under different criteria for stable mass transfer (see Sect.~\ref{sec:criteria}). The solid black line shows the critical mass ratios resulting from our quasi-adiabatic criterion, with the grey shading indicating the parameter space where mass transfer would be unstable. The red lines show the critical mass ratios resulting from the dynamical evolution criterion with $A_{\rm dyn}~= 0.01, 0.05, 0.25$, as indicated to the right end of the lines. The solid orange line shows the critical mass ratios resulting from the OLOF-based criterion, and the solid blue line and hatching show where the Darwin instability could be encountered (see Sect. \ref{sec:discussion_limitations}). The solid horizontal grey line shows where $M_{\rm a}/M_{\rm d} = 1$, and the dashed vertical grey lines show the end of the HG phase (panel a and left line in panel b) and the tip of the RGB (panel b). } 
     \label{fig:criteria_compare_qcrits_1n4}
\end{figure}

\begin{figure*}
\centering
\includegraphics[width=\hsize]{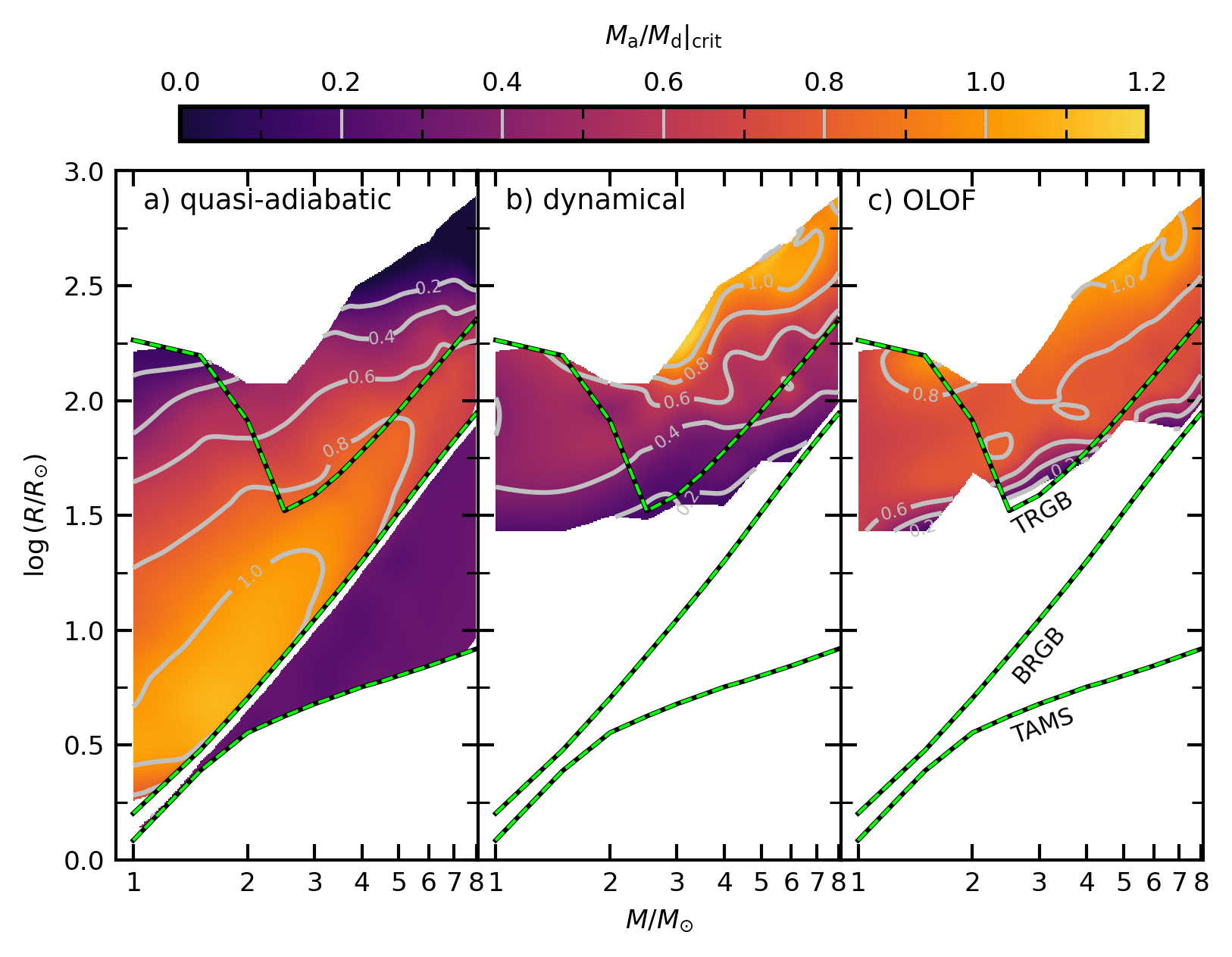}
  \caption{Critical mass ratios ($q_{\rm crit}$, where $q\equiv M_{\rm a}/M_{\rm d}$) for all our binary models as interpolated contours. The three panels show critical mass ratios derived using different criteria (outlined in Sect. \ref{sec:criteria}), as indicated at the top of each panel. The colour shows the value of the critical mass ratio, following the colour bar at the top of the figure. The results shown in panel b are for $A_{\rm dyn} = 0.05$. The green dashed lines show the donor radii corresponding to the end of the main sequence (TAMS), the base of the RGB (BRGB), and the tip of the RGB (TRGB), from bottom to top.}
     \label{fig:qcrit_main}
\end{figure*}

In this section we present our results for the critical mass ratios introduced in Sec.~\ref{sec:criteria}. For mass ratios greater than these critical ratios ($M_{\rm a}/ M_{\rm d} > q_{\rm crit}$), mass transfer was found to be stable according to one of the three criteria we applied. In Fig.\,\ref{fig:criteria_compare_qcrits_1n4}, we present the critical mass ratios resulting from these criteria for two representative ZAMS masses $1~M_\odot$ and $4~M_\odot$. 

For HG donor stars, we find that any mass transfer instabilities set in well before their envelopes have become even marginally convective. Hence, the response of these stars to mass loss is essentially that of a donor star with a completely radiative envelope. Therefore, mass transfer from HG donors is relatively stable (see Sect. \ref{sec:crit_th}), with the typical quasi-adiabatic critical mass ratio being around $q_{\rm qad}=0.25$, which agrees well with values typically assumed for such donors \citep[see e.g.][]{Cla14}. Our $1\,M_\odot$ model with the smallest Roche radius, which is classified as HG, behaves more like a convective donor star (discussed in more detail below), due to the presence of a non-negligible surface convection zone. As a result, this subgiant donor has comparatively larger $q_{\rm qad}$ than the HG donor stars discussed above. We were unable to determine $q_{\rm dyn}$ and $q_{\rm OLOF}$ for any HG donor. In our simulations, the mass-transfer rate and corresponding orbital evolution of the system never reach the thresholds for dynamical mass transfer, as described in Sect. \ref{sec:crit_dyn}. Similarly, our model HG stars never expand enough relative to their Roche lobes to overflow their OL in these systems.

Also visible in Fig.\,\ref{fig:criteria_compare_qcrits_1n4}b is the sharp transition between the HG stars and giants, where the critical mass ratio abruptly increases by a factor of roughly 3-4. This coincides with the region where the structure of the outer layers of the donor stars changes significantly as a function of stellar radius. A significant part of the envelope of giant stars is already convective prior to the onset of mass transfer. As discussed in Sect.~\ref{sec:results_mdot}, stars with convective outer layers tend to expand in their initial response to mass loss \citep[note that even for polytropic models this behaviour depends on the fractional core mass; see][]{Hje87} and hence transfer mass occurs at higher rates compared to stars with radiative envelopes. This results in significantly larger values for the critical mass ratio (i.e.\ less stable mass transfer) for the least evolved giants compared to the most evolved HG stars.

For giant stars, there is no visible discontinuity in $q_{\rm qad}$ values between donor stars filling their Roche lobes on the RGB, and those that do so on the AGB. Whilst there are structural differences between stars in both phases, these are mainly in the star's core and burning shell(s). The outer layers are very similar in structure and deeply convective on both giant branches, and hence the trends in response to mass loss are very similar. As donor stars become more evolved giants, their core mass fraction increases (except for dredge-up events) and the local thermal timescales in the outer layers decrease (raising $\dot{M}_{\rm th,crit}$). Both have a stabilising effect on the mass transfer (see Sect. \ref{sec:crit_th}) and $q_{\rm qad}$ decreases with increasing donor radius. For those donor stars that fill their Roche lobes near the thermally pulsing part of the AGB, the local thermal timescales in the outer layers relevant for the response to mass transfer are short enough that $q_{\rm qad}$ can be even smaller than for HG donor stars. 

With increasing donor radius on the RGB and AGB, the dynamical timescale decreases and the mass-transfer rates increase, while the donor stars extend further beyond their Roche lobes. Therefore, both $q_{\rm dyn}$ and $q_{\rm OLOF}$ typically follow a trend opposite to $q_{\rm qad}$; these critical mass ratios generally increase with both $\log R$ and $M_{\rm d}$. In the systems with giant donors for which we could not determine the critical mass ratios $q_{\rm dyn}$ and $q_{\rm OLOF}$, the mass transfer turned unstable and our MESA simulations ended before the corresponding criteria could be reached, under conditions in accordance with our quasi-adiabatic criterion (see Sect. \ref{sec:crit_th}).

For our choice of $A_{\rm dyn} = 0.05$, the dynamical and OLOF-based criteria result in similar critical mass ratios at the largest radii ($\log R/R_\odot \ga 1.5$). At these initial donor radii, both of these criteria are typically met during the mass transfer in binaries where $M_{\rm d} \geq 2.5\;M_\odot$. We stress again, as discussed in Sect. \ref{sec:results_mdot} that in this regime (where typically $q_{\rm qad} < q < \{q_{\rm dyn}, q_{\rm OLOF}\}$) we do not find the mass transfer to run away and become unstable. Instead, the mass transfer is self-regulating and proceeds in a stable manner down to $q = q_{\rm qad}$. However, in this part of the parameter space, multiple assumptions in our calculations (such as the Roche geometry and spherical symmetry) break down, and our results start to become unreliable. As such, the OLOF and dynamical criteria should be considered as indicators of where 1D-based calculations break down, rather than criteria for when mass transfer turns unstable.

Also shown in Fig.\,\ref{fig:criteria_compare_qcrits_1n4} is the effect of choosing different values for $A_{\rm dyn}$ (see Eq. \ref{eq:dynev_crit}). The choice of $A_{\rm dyn}$ is somewhat arbitrary, as, to the best of our knowledge, there does not exist an obviously most suitable value for it. Adopting a larger value of $A_{\rm dyn}$ means that the evolution of binaries must become more rapid in order to be considered dynamical. This happens when the mass ratios are more extreme, such that larger values of $A_{\rm dyn}$ lead to lower critical mass ratios.

We summarise our results for the critical mass ratios for the full parameter space for all three stability criteria in Fig.\,\ref{fig:qcrit_main}. In general, the critical mass ratios depend on the mass and radius of the donor star when it fills its Roche lobe. However, the critical mass ratios have comparable values for donor stars in similar evolutionary stages. This holds especially well for the low-mass ($M \leq 2~M_\odot$) stars in our simulations. As described in more detail above, the critical mass ratios vary with stellar radius and mass in systematic manners: $q_{\rm qad}$ is roughly constant for HG stars, increases by a factor of 3-4 at the transition to the giant phase, during which $q_{\rm qad}$ typically decreases with $\log R_{\rm d}$. The maximum value of $q_{\rm qad}$ generally decreases with $M_{\rm d}$. The other critical mass ratios, $q_{\rm dyn}$ and $q_{\rm OLOF}$ follow the opposite trend and tend to increase with $\log R_{\rm d}$ and $M_{\rm d}$. For most of the parameter space studied in this work, we find that $q_{\rm qad}$ is smaller than 1, except for a small region where the maximum critical mass ratio we find is $1.05$.

\section{Discussion}
\label{sec:discussion}
In this section we compare our results to prior literature, and to commonly used critical mass ratios in BPS works. Furthermore, we critically examine sources of uncertainty in our results, and discuss several simplifying assumptions we made. We end by outlining the implications of our results for CE evolution, and a comparison of our results to observations.

\subsection{Comparison to previous works}
\begin{figure*}
\centering
\includegraphics[width=\hsize]{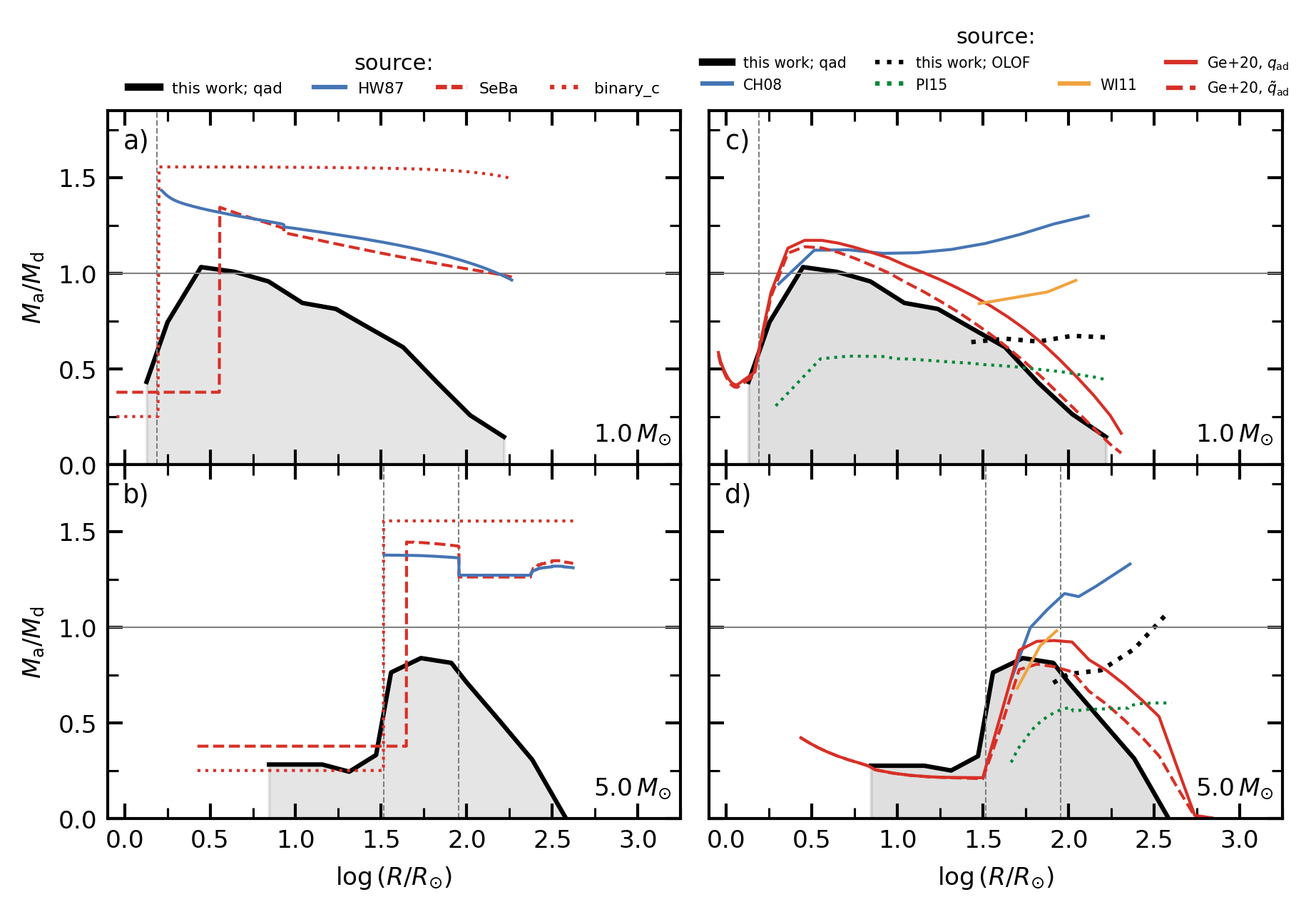}
  \caption{Comparison of our results to the default sets of critical mass ratios ($M_{\rm a}/M_{\rm d}$) implemented in the \texttt{SeBa} and \texttt{binary\_c} rapid BPS codes and those found in detailed theoretical modelling efforts as a function of donor star radius. Each line corresponds to the source with the same style in the legend. As in Fig.~\ref{fig:criteria_compare_qcrits_1n4}, the solid black line and grey shading indicate the part of the parameter space that would lead to unstable mass transfer according to our quasi-adiabatic criterion.}
     \label{fig:qcrit_compare_literature}
\end{figure*}

\label{sec:discussion_comparison}

The classic work of \citet[][hereafter HW87]{Hje87} provides systematic values for the critical mass ratio and adiabatic mass-radius exponent $\zeta_{\rm ad}$ for a wide parameter space. The authors approximated the structure of giant stars as condensed polytropes that respond to mass loss adiabatically. By comparing the response of the stellar radii to the response of the Roche lobe, the authors derived critical mass ratios for dynamically unstable mass transfer as a function of the core mass fraction. We fitted their results (provided in Table 3 of \citetalias{Hje87}) by the following function, which is accurate to within $0.5\%$ over the full range: 
\begin{align}
    q_{cr, HW87} = 1.577 \cdot \frac{(1-m_{\rm c}^{3.533})^{0.924}}{1 + m_{\rm c}^{0.929}},
\end{align}
where $m_{\rm c}$ denotes the core mass fraction and $q = M_{\rm a}/M_{\rm d}$, as before. We assume the core mass fraction to be the same function of initial donor mass and radius as in our single-star models, and compare $q_{\rm cr, HW87}$ to our results in Fig. \ref{fig:qcrit_compare_literature} panels a) and b). We find that our critical mass ratios are significantly lower over the full parameter range studied in this work (i.e.~we find mass transfer to be more stable than predicted by \citetalias{Hje87}), in qualitative agreement with earlier work. The difference increases with the radius at which the donors fill their Roche lobes. Our calculations do not assume that the donor star responds adiabatically and we find that our donor stars can find a state of self-regulating mass transfer down to significantly lower mass ratios. Whereas virtually all critical mass ratios of \citetalias{Hje87} for the parameter space studied here are above unity, our results indicate that $q_{\rm qad} < 1$ (see Sect. \ref{sec:results_qcrit}).

Rapid BPS codes typically rely on simplified prescriptions for either the critical mass ratios or $\zeta_{\rm ad}$. The resulting boundaries for stable mass transfer are a collection of several literature values and fits, which are not necessarily consistent with one another in terms of the assumptions made in their derivation. We compare our results to the prescriptions for the critical mass ratios used in the \texttt{binary\_c} code \citep[see][and references therein]{Izz04,Izz18} as given in Table 2 of \citet{Cla14}. These critical mass ratios are based on \citet{Hur02}, who use approximate fits to single stellar models to obtain $\zeta_{\rm eq}$ (the response of a donor star's equilibrium radius)  for giants, and assume that $\zeta_{\rm ad} = \zeta_{\rm eq}$. This is the default prescription in  \texttt{binary\_c}, but we note that other options are available to users. Additionally, we compare our results to the prescriptions for the $\zeta_{\rm ad}$ values used in the \texttt{SeBa} code \citep[][as listed in Table A.1 of the latter]{Por96,Too12}. The \texttt{SeBa} code relies on the results from \citetalias{Hje87} for giants with deeply convective envelopes, and employs a constant $\zeta_{\rm ad} = 4$ for HG stars, as well as giants with shallow convective envelopes. As before, we use the core mass fraction and donor mass as a function of donor radius from our own single-star simulations. We convert the aforementioned expressions for $\zeta_{\rm ad}$ to critical mass ratios using the relations presented in \citet{So97}, under the assumption of conservative mass transfer, appropriate for comparison to our results.

For HG donor stars, the critical mass ratios of \texttt{binary\_c} are similar to our results, whilst those from \texttt{SeBa} are systematically larger, as can be seen in Fig. \ref{fig:qcrit_compare_literature} panels a) and b). The transition from radiative to convective stars occurs at different radii between the two BPS codes and our models. In part, this is because \texttt{SeBa} employs a different prescription to determine beyond which radius the giants respond as convective stars during mass loss. For the giant stars, the critical mass ratios are lower in \texttt{SeBa} than in \texttt{binary\_c}. This difference results from the assumption made in \texttt{binary\_c} that $\zeta_{\rm ad} = \zeta_{\rm eq}$, whereas in reality $\zeta_{\rm ad} > \zeta_{\rm eq}$. Both sets of values, however, are significantly larger than ours, by a factor of $\sim$ 2--3. Whereas we find a strong decrease in the critical mass ratio along the giant branches, such a decrease with radius is much milder or absent in both BPS codes (except for the occasional jump due to dredge-ups). This suggests that predictions based on these popular rapid BPS models not only assume mass transfer to be less stable than we conclude, but also miss an important systematic effect in mass-transfer stability as a function of the donor radius at the onset of RLOF.

We next compare to results from various model calculations that do not implicitly assume an adiabatic response of the donor. By computing binary evolution models with the \texttt{STARS/ev} code \citep[][]{Egg71,Pol95, Gleb08}, \citet{Che08} (CH08) find the critical mass ratios for a similar parameter space as ours. Their results for giant stars exhibit a trend mostly opposite to our quasi-adiabatic critical mass ratios, more akin to our OLOF and dynamical criteria, and are typically larger than $q_{\rm qad}$ (see Fig. \ref{fig:qcrit_compare_literature} panels c) and d)). This might be due to the different way the mass-transfer rate is computed in their evolution code, which we discuss in more detail in Sect. \ref{sec:discussion_mtransfer_scheme}. For their most evolved donor stars, the critical mass ratios found by \citetalias{Che08} exceed those from \citetalias{Hje87} and those used in \texttt{SeBa}. 

The same \texttt{STARS/ev} code, along with a `standard Henyey-type code' \citep[more details in][]{Iva04} was used by \citet{Woo11} (WI11) to perform a similar analysis. These authors focused on demonstrating the importance of the superadiabatic surface layer for the response of giant donors to mass loss and, as such, the parameter space they study is limited. The critical mass ratios they find follow a trend similar to those found by \citetalias{Che08}, but offset to lower mass ratios. This offset could be explained by the difference in choice for the free parameter $C$ in the expression for the mass-transfer rate ($C=1$ in WI11 whilst $C=500$ in \citetalias{Che08}; see these works for more details), which according to the authors affects mass-transfer stability (but see Sect. \ref{sec:discussion_mtransfer_scheme}). For low-mass stars, their critical mass ratios are significantly larger than the ones presented in this work, but they are more comparable for intermediate-mass stars, at least in the limited parameter space for which they were computed.

\citet[][hereafter PI15]{Pav15} studied the response of giant stars to mass loss in a similar manner as this work. These authors used a modified version of MESA for their simulations, with a set-up and physical parameters comparable to our simulations. The main differences between their simulations and ours are their inclusion of hydrodynamic terms in the stellar structure and evolution equations, and their modifications to the \cite{Kol90} mass-transfer scheme (see Sect. \ref{sec:discussion_mtransfer_scheme}). We compare our results to the critical mass ratios of \citet{Pav15} as approximated by their `$L_2/L_3$-overflow condensed polytrope simplification' results, which we extracted from their Fig. 15. Because their criterion for unstable mass transfer is based on $L_2/L_3$-overflow, we would expect it to be quantitatively similar to our OLOF criterion (see Sect. \ref{sec:crit_ol}). Our values for $q_{\rm OLOF}$ follow a similar trend to the critical mass ratios of \cite{Pav15}, but with a substantial offset to larger values by $\Delta ({M_{\rm a}}/{M_{\rm d}}) \sim 0.15$ or more.

However, as can be observed in Fig. 14 of PI15, their $L_2/L_3$-overflow condensed polytrope simplification estimate of the critical mass ratios is not always consistent with their detailed calculations. 
In the case of a $1\,M_\odot, 100\,R_\odot$ giant, our critical mass ratio for OLOF agrees very well with their detailed result, even though the difference between our result and their simplified approximation is significant. For this specific case their calculations for the mass-transfer rate and amount of overflow (as shown in their Fig. 12) also agree very well with ours. On the other hand, for less evolved $1\,M_\odot$ donor stars, their simplified approximation is closer to the results from their detailed calculations. Likewise, for a $2\,M_\odot$ donor star shown in their Fig. 14, the difference between their detailed calculations and their simplified approximation depends on the donor radius. Hence, the difference between our results and theirs cannot always be explained by this simplification. Lastly, these authors do not appear to encounter the limitation set by the critical thermal rate, although it is unclear why.

In a series of works focusing on the systematic study of mass-losing stars in binaries, which are similar in spirit to \citetalias{Hje87} but do not rely on polytropic approximations, \cite{Ge10,Ge15,Ge20a} infer critical mass ratios from simulations of single stars. Under the assumption that the donor stars respond to mass loss adiabatically, the critical mass ratios are determined as the smallest mass ratio for which the mass-transfer rate does not exceed the global KH limit. They calculate the mass-transfer rate using a prescription very similar to to our Kolb scheme. The resulting critical mass ratios ($q_{\rm ad}$ in \citealp{Ge10,Ge15,Ge20a}) are very similar to our $q_{\rm qad}$ for HG stars, but different for giants. For giant donors, both sets of results follow a similar trend, but our critical mass ratios are systematically smaller by an almost constant difference $\Delta ({M_{\rm a}}/{M_{\rm d}}) \sim 0.2$, meaning we typically find mass transfer to be more stable. Since we do not impose an adiabatic response on our donor stars, mass transfer can remain stable even when the mass-transfer rate greatly exceeds the global KH rate, as shown in Sect. \ref{sec:results_mdot}. Instead, we find that the limiting mass-transfer rate is best taken to be one set by the shortest local thermal timescale in the envelope's superadiabatic layers. To investigate whether this could be contributing to the difference between our results and those of \cite{Ge10,Ge15,Ge20a}, we re-define our critical mass ratios in a similar manner to theirs; by setting the limiting mass-transfer rate to the global KH rate instead. We find that this leads to very similar critical mass ratios for low-mass stars ($M_{\rm d}\leq 2~M_\odot$), although our critical mass ratios deviate more from their results for larger initial donor masses.

In addition to their models with a realistic envelope structure, \cite{Ge10,Ge15,Ge20a} constructed adiabatic mass loss sequences involving stars with artificially isentropic envelopes, thus neglecting the superadiabatic layers. This is intended to mimic the effect of thermal relaxation of the surface layers (as shown in Fig. 2 of \citetalias{Woo11} and our Fig. \ref{fig:entropy_mass_transfer_response}). The resulting critical mass ratios based on these isentropic envelope models ($\tilde{q}_{\rm ad}$; the red dashed lines in Fig. \ref{fig:qcrit_compare_literature} c and d) are much more similar to our set of $q_{\rm qad}$ values than the \cite{Ge20a} values based on more realistic envelopes ($q_{\rm ad}$, red solid lines in Fig. \ref{fig:qcrit_compare_literature} c and d). Apparently the effects of the two approximations made by Ge et al  (neglect of thermal relaxation in the outer layers, and neglect of the superadiabatic envelope structure) nearly compensate each other to give critical mass ratios very similar to our more realistic assumptions. We also note that, in Fig. \ref{fig:qcrit_compare_literature}d, the results of \cite{Ge10,Ge15,Ge20a} (both ${q}_{\rm ad}$ and $\tilde{q}_{\rm ad}$) are shifted to somewhat larger radii compared to our $q_{\rm qad}$ values. This is probably due to differences in the physical assumptions made in the calculations regarding the amount of overshooting, the adopted mixing length and/or wind mass loss.

\subsection{Limitations of the model}

\label{sec:discussion_limitations}
Here, we outline the most important simplifications we made in our calculations, and discuss how varying these might affect our results.

\subsubsection{Treatment of the accretor and mass-transfer efficiency}

We have ignored the evolution of the accretor in this work, treating it as an all-accreting point mass. However, at sufficiently high mass-transfer rates, accreting stars are unable to readjust their structure on timescales shorter than the mass transfer timescale, such that the accretor will be significantly brought out of thermal equilibrium. For those accreting main-sequence stars that have radiative envelopes ($M_{\rm a} \ga 1.5~M_\odot$), this causes substantial expansion of the accretor star beyond its main-sequence radius \citep{Kip77,Neo77,Pac79}. As a result, the secondary may fill its own Roche lobe, leading to the formation of a contact binary \citep{Pol94,Wel01}. Although such a contact configuration does not necessarily lead to a CE, it could lead to OLOF from the accretor, and thus to non-conservative mass transfer \citep{Mar16}. Nonetheless, some stellar systems have been explained as a consequence of orbital instability after the accreting star expands into contact, including some apparently post-merger stars  \citep[e.g.][]{LangerHeger1998,Justham+2014}, and close hot-subdwarf binaries \citep{Justham+2011}.

Such a contact configuration may be avoided if stellar rotation is included. During mass transfer, the accreting star not only gains mass, but also angular momentum, which causes it to rotate faster. As the accretor spins up, the accretor's surface rotation rate can eventually exceed the rotational break-up rate. It has been demonstrated \citep{Pac81} that very little mass is required to spin up an accreting star to this break-up speed. Binary evolution models in which this effect is accounted for \citep{Pet05,DeM13} predict almost fully non-conservative mass transfer in massive binaries \citep[however, also see][for a different view]{DeC13}. Additionally, rapid accretion could potentially produce strong winds and/or jets from an accreting main-sequence star \citep[as seen in some binary post-AGB stars; see e.g.][]{Win18}

Observational evidence supports a wide range of mass-transfer efficiencies, from almost conservative to almost fully non-conservative \citep[see e.g. the references above and][]{DeG94,Nel01,DeM07,Sch18,Vin20}. Additionally, different forms of non-conservative mass transfer would have very different associated AM loss, which would affect the orbital evolution of the binaries. These remain as-of-yet unsolved problems in the evolution of interacting binary systems. Non-conservativeness during mass transfer, whether for the reasons just described or otherwise, affects the orbital evolution. In this work, we have assumed all of the mass transferred from the donor to be accreted by its companion. As mentioned in Sect.~\ref{sec:method}, under the often-made assumption that mass leaving the system carries with it the specific orbital angular momentum of the accreting star, the fully conservative case is expected to be the least stable one, based on orbital evolution arguments. As such, our results would provide an upper limit on the critical mass ratios separating stable and unstable mass transfer. We investigated the effects of non-conservative mass transfer by re-simulating a selection of systems representative of the full span of our parameter range with different (constant) mass-transfer efficiencies. We find that our completely conservative mass transfer calculations indeed provide a lower limit on the stability boundary.

\subsubsection{Tidal interactions}

Tidal interactions can act to affect the orbital evolution and therefore, potentially, the stability of mass transfer. For example, in close binaries, tidal interaction with the companion can transfer spin angular momentum back into the orbit, possibly preventing critical rotation. The effect of spin-orbit coupling on mass transfer was studied in detail by \cite{Mis2020}, in the context of low- and intermediate-mass X-ray binaries containing neutron star accretors. They found that the inclusion of tidal interactions only had a minor effect on the stability of mass transfer. Since, in most of our systems, the main-sequence accretors are relatively compact compared to their post-MS companions, we conclude that the inclusion of spin-orbit coupling would not significantly alter our results or main conclusions. However, at sufficiently small mass ratios, tidal spin up of the donor star can lead to an ever growing orbital angular frequency, decreasing the orbital separation in a spiral-in manner. This phenomenon, known as the Darwin instability \citep[see e.g.][]{Dar1879, Cou73, Hut80}, occurs when the spin angular momentum of one of the binary components exceeds one third of the binary orbital angular momentum. Our calculations do not take this instability into account, but we estimate in which part of the parameter space it is expected to occur and where tidal effects may affect our results significantly. To that end, the donor radius at the moment it fills its Roche lobe should not exceed
\begin{align}
    \frac{R_{\rm d}}{a} = \left[3 r_{\rm g}^{2} (1 + q^{-1}) \right]^{-\frac{1}{2}}
\end{align}
in order to avoid a Darwin instability \citep[e.g.][]{Dar1879, Egg01, Mac17}. Here, $r_{\rm g}$ is the donor stars' dimensionless radius of gyration ($r_{\rm g}^2$ is defined as its moment of inertia $I = \frac{8\pi}{3} \int_0^{M_{\rm d}} \rho(r)r^4 \text{d}r$, divided by $M_{\rm d}R_{\rm d}^2$). 
We equate this expression to Eq. \ref{eq:rL1}, and solve for $q$ using our single-star models that form the initial configuration of our binary simulations. The resulting critical mass ratio for the Darwin instability is shown as a blue line in Fig.~\ref{fig:criteria_compare_qcrits_1n4}. As can be seen in that figure, the Darwin instability occurs only for mass ratios below our critical values for stable mass transfer, except perhaps in the most evolved intermediate-mass giants. Therefore, we do not expect the inclusion of tidal affects would alter our main results and conclusions.

\subsubsection{The mass-transfer scheme}
\label{sec:discussion_mtransfer_scheme}
The comparison of our results to the literature in Sec.~\ref{sec:discussion_comparison} suggests that the method used to calculate the rate of mass transfer can have strong effects on the stability of the RLOF, at least for evolved giant donors. In this work we used the prescription given in \cite{Kol90} as implemented in MESA \citep{Pax15} (the Kolb scheme). This method treats mass transfer as resulting from a laminar, subsonic flow along equipotential surfaces in the layers of the donor star that exceed the Roche lobe, towards the nozzle around the $L_1$ point. For the high mass-transfer rates that are relevant here, these layers are optically thick and the flow is assumed to be adiabatic with a constant adiabatic index. The vertical stratification of these layers is assumed to be hydrostatic and taken from the 1D stellar (MESA) model, so that the method explicitly takes into account the density structure in the overflowing layers. 

By contrast, in some of the literature we compared with (in particular the studies of \citetalias{Che08} and \citetalias{Woo11}) the mass-transfer rate is taken to be a function of only the over-extension of the photospheric radius of the donor star beyond its Roche radius (i.e. $\dot{M}_{\rm d} \propto [(R_{\rm d} - R_{\rm L})/R_{\rm L}]^3$). Such a relation is expected if the outer layers have a polytropic structure with index $n=1.5$ \citep{Pac72}, which would be appropriate for an adiabatically stratified convective envelope. However, the proportionality factor is taken to be a constant free parameter ($C$) rather than calculated from the physical properties of the outer layers. This approach is sufficient for relatively unevolved donor stars with mass transfer on a thermal or nuclear timescale, where the resulting mass-transfer rate is insensitive to the precise method of computation \citep{Kol90}. On the other hand, for evolved red giants this simple approach can be problematic, for two reasons. Firstly, the outer envelopes of an extended convective giant are not adiabatically stratified but substantially superadiabatic. Secondly, the layers close to the photosphere have very low density and a pressure scale height that can be a substantial fraction of the stellar radius. This means that these stars must typically extend significantly past their Roche lobes for mass to be transferred at or above the KH rate. This is not taken into account in the simplified prescription used by for example, \citetalias{Che08} and \citetalias{Woo11}, which could therefore yield mass-transfer rates that are too high. This could explain why such studies result in larger critical mass ratios than our results, and why the difference grows with increasing radius (Fig. \ref{fig:qcrit_compare_literature}c and d). Conversely, the similar trend of critical mass ratio with stellar radius found by \cite{Ge20a} could result from their use of a method to compute the mass-transfer rate very similar to our Kolb scheme.

More recent work by \cite{Pav15} and \cite{Marchant+2021} improved the \cite{Kol90}  method for calculating the mass-transfer rate by taking into account the detailed shape of the Roche potential around $L_1$ and not approximating overflowing optically thick layers as an ideal gas, computing $P/\rho$ from their detailed stellar models instead. These modifications are not expected to significantly change our results, as ideal-gas pressure always dominates in the outer layers of a red giant, and the effect of the improved $L_1$ geometry is small for mass ratios that are not too extreme (see Appendix~A of \citealp{Marchant+2021}). Indeed, for a $1~M_\odot$, $100~R_\odot$ giant donor we find almost the same mass-transfer rate and RLOF factor as \cite{Pav15} (their Fig.~12).

Nevertheless, we should consider that other assumptions made in the \cite{Kol90} method may break down for evolved, very extended giants transferring mass at high rates. The assumed laminar flow may become turbulent and the structure of the overflowing layers may no longer be in hydrostatic equilibrium, even at locations far away from $L_1$. A 3D hydrodynamical treatment of the flow may be required to compute the mass-transfer rate in this case, and may lead to a different relation between $\dot M$ and the degree of Roche-lobe overfilling, potentially affecting the resulting critical mass ratios for all three criteria we consider in this paper.

\subsubsection{Convection}

Lastly, an appropriate description of convection is essential to determine the structure of a red giant. A reliable treatment of superadiabatic convection would require 3D hydrodynamical and radiation transport calculations \citep[e.g.][]{Freytag+Salaris1999,Trampedach+2014,Magic+2015,Gol22}. A widely adopted solution within current 1D stellar evolution codes is different variants of the mixing-length theory \citep[MLT;][]{Boh58}, a simplified analytical formulation of the problem. Whilst effective and widely in use, the MLT might not always be an appropriate treatment, especially in the scenarios considered here. Even for single-star models, studies such as the aforementioned \citet{Gol22} show that the more realistic near-surface structure of giants can differ from those calculated using the MLT framework. Moreover, MLT implicitly assumes that the convective region is in a steady state, or at least that the timescale of change is much longer than the convective turnover time. In our simulations, as the evolution of the donor approaches the dynamical timescale, the timescale of the evolution becomes similar to or shorter than the convective turnover time when subsonic convective velocities are assumed. In this regime the approximations in MLT might be so inappropriate as to diminish the validity of our results.

\subsection{Implications for common envelope evolution}
\label{sec:implications_CE}

In this work we have studied mass transfer in low- and intermediate-mass binaries and calculated three different critical mass ratios for stable mass transfer, corresponding to the three criteria described in Sect. \ref{sec:criteria}. For systems with $q < q_{\rm qad}$, unstable, runaway mass transfer appears to be guaranteed and will almost inevitably lead to a CE. This is broadly consistent with observational constraints from post-CE binaries (see Sect. \ref{sec:discussion_observations}).

However, our models cannot always provide a definite prediction of the outcome of mass transfer in binaries with $q > q_{\rm qad}$. This is partly because of the limitations discussed in Sect. \ref{sec:discussion_limitations}. The outcome of mass transfer in systems with $q_{\rm qad} < q < \{q_{\rm dyn}, q_{\rm OLOF}\}$ is particularly uncertain. In our models we find self-regulated mass transfer in this regime, albeit at very high rates, and we were able to simulate the entire mass transfer for most of these systems. One has to keep in mind, however, that we did not simulate OLOF and its effects on the orbital evolution of the binaries in this work, and the MESA code (in part due to its 1D nature) is not designed to simulate the types of rapid binary interactions that occur in systems with $q \la q_{\rm dyn}$. Additionally, and perhaps more importantly, the outcome of mass transfer in such cases will crucially depend on the reaction of the accretor, which we did not take into account. The very high mass-transfer rates suggest that very little transferred material can be directly accreted by the companion star, and the excess material may rapidly fill the Roche lobe of the companion. Whether this leads to the onset of a CE followed by a spiral-in or to some other, less dramatic form of non-conservative mass transfer cannot be predicted from our models.

Finally, systems that can be classified as stable by all three of our criteria have the best prospects of avoiding a CE altogether. This by itself would significantly enlarge the parameter space for stable mass transfer compared to what is currently assumed in many popular rapid BPS codes.

\subsection{Comparison with observed systems}
\label{sec:discussion_observations}

\begin{figure*}[h!]
\centering
\includegraphics[width=\hsize]{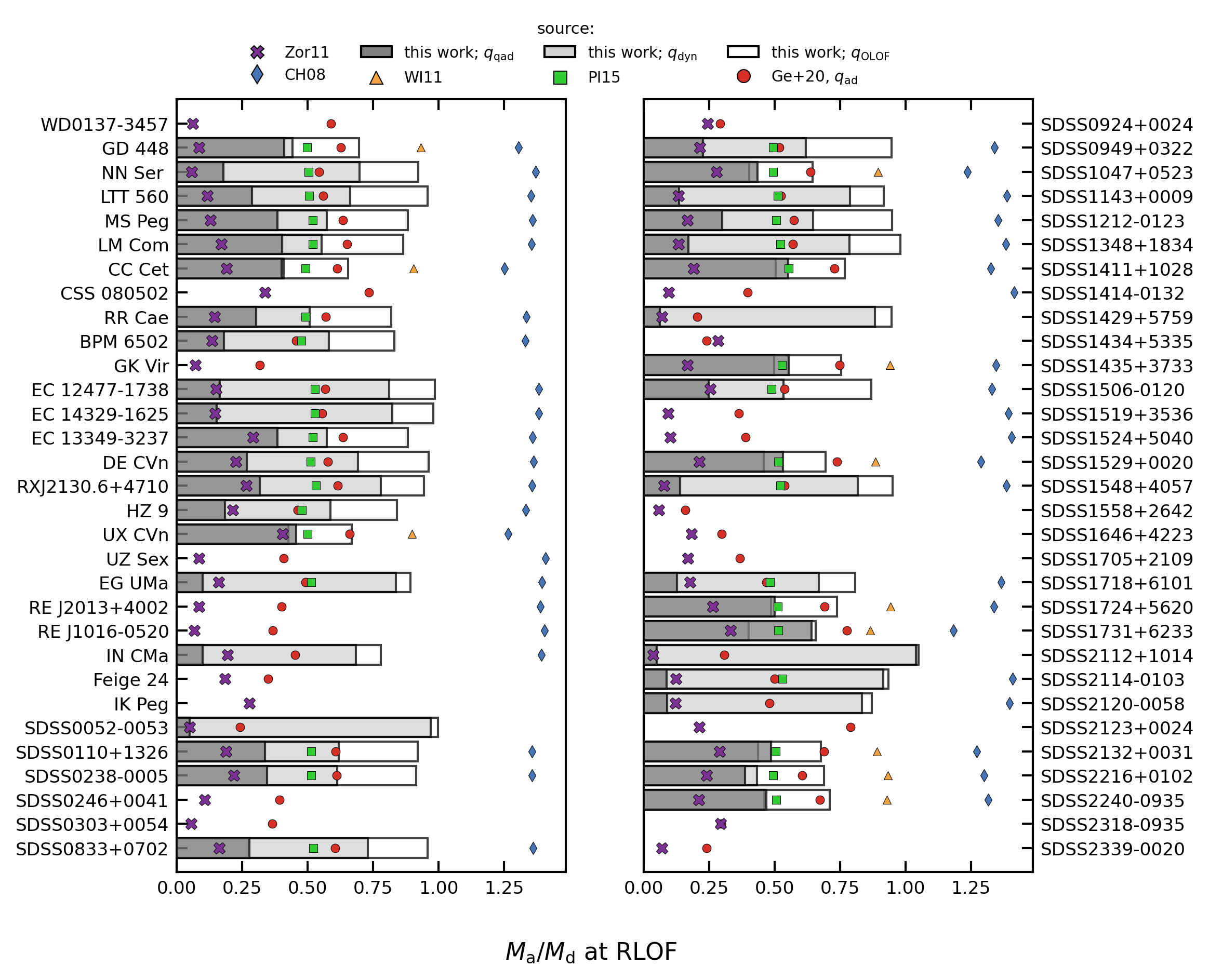}
  \caption{Comparison of our results to observationally derived mass ratios. The purple `x' markers indicate mass ratios of systems at the onset of mass transfer, inferred from observations of post-CE binaries. Different shades of grey show mass ratio ranges leading to unstable mass transfer based on the different criteria studied in this work, as indicated in the legend, interpolated to the properties of each system. The coloured symbols correspond to interpolated critical mass ratios based on results from a variety of sources, as indicated in the legend using the same abbreviations for literature sources as in the text and as used in Fig. \ref{fig:qcrit_compare_literature}.}
     \label{fig:qcrit_compare_obs_and_lit}
\end{figure*}

To check that our results are consistent with expectations from observations, we compare our theoretical predictions to the sample of 62 post-common-envelope binaries (PCEBs) constructed by \cite{Zor11}. The pre-CE properties of the binaries were inferred by the authors as follows. For each binary in the sample, the mass of the white dwarf was assumed to be equal to the pre-CE core mass of the donor star, and the mass of the companion star was assumed to be unchanged by the CE event. The single-star evolution code from \cite{Hur00} and the envelope binding energy prescription from \cite{Zor10}, where the energy used to unbind the envelope is assumed to be a fraction $\alpha$ of the change in orbital energy during the spiral-in, were employed to calculate the pre-CE mass and radius of the donor star. These were then used to calculate the mass ratios at the onset of the ultimately unstable mass transfer ($q_{\rm obs}$), which are shown in Fig. \ref{fig:qcrit_compare_obs_and_lit} with purple crosses. We note that the assumption made by \cite{Zor11} is that all the systems in their catalogue indeed experienced a CE. For IK Peg, we are not convinced this is indeed the most plausible formation scenario, given its wide current period of 21.7~days.

Also shown in Fig.~\ref{fig:qcrit_compare_obs_and_lit} by dark grey shading are our theoretical predictions for the range of mass ratios that lead to a CE, based on our $q_{\rm qad}$ for the assumed pre-CE donor mass and radius. The typical inferred progenitor of these systems is an evolved giant, in some cases on the thermally pulsing phase on the asymptotic giant branch  (TPAGB), which means that we cannot always compare their inferred properties to our predictions, because we do not model mass transfer that begins during the TPAGB phase of our donor stars. Of those observed PCEBs for which comparison to our predictions is possible, we find that 33 out of 41 systems (roughly 80\%) have $q_{\rm obs} \leq q_{\rm qad}$, that is, they are consistent with our predictions for when a CE is unavoidable. However, only for two of the eight systems with $q_{\rm obs} > q_{\rm qad}$ (EG UMa and IN CMa) we find this difference to be significant. Because uncertainties on the inferred pre-RLOF mass ratios are not provided in \cite{Zor11}, we estimated them in the following manner. We use the provided uncertainties on the observed white-dwarf mass and companion mass, as well as the range of values for the $\alpha$ parameter that \cite{Zor10} claim can reproduce all the systems in this sample: $0.2\leq \alpha \leq0.3$ (though we note that for each individual system the range of $\alpha$ values the authors allow is much larger). Then, varying these parameters over their respective ranges of uncertainty individually and independently, we solved Eqs. 3-6 in \cite{Zor10} to derive an estimate of the uncertainty on $q_{\rm obs}$. For six of the eight systems with $q_{\rm obs} > q_{\rm qad}$ the difference with our results is within this uncertainty. All eight systems furthermore have $q_{\rm obs} \leq \{q_{\rm OLOF}, q_{\rm dyn}\}$, meaning that these systems are all consistent with stability limits derived from those criteria. However, as we stress in Sect.~\ref{sec:implications_CE}, this by itself does not necessarily mean a CE is unavoidable based on our results. All things considered, our predictions for the stability of mass transfer agree reasonably well with the inferred properties of PCEBs.

Figure~\ref{fig:qcrit_compare_obs_and_lit} also shows critical mass ratios from literature, similar to Fig.~\ref{fig:qcrit_compare_literature}. As it can be seen, the inferred pre-CE mass ratios are in all cases smaller than the critical mass ratios derived in these previous studies. This is not surprising given that our $q_{\rm qad}$ values tend to be smaller than those from the literature.

\section{Conclusions}
\label{sec:conclusion}

We have performed a systematic theoretical investigation of the critical mass ratios for stable mass transfer in low- and intermediate-mass binaries. We used the MESA stellar evolution code to simulate the mass-transfer evolution of 1,404 binaries, with post-MS donor stars ranging from $1~M_\odot$ up to and including $8~M_\odot$ and with a wide range of mass ratios and orbital periods. 

We explored three different criteria for identifying unstable mass transfer. Two are well-known and commonly used criteria, based on dynamical-timescale evolution and OL overflow by the donor star. Additionally, we used a criterion based on the transition to an effectively adiabatic response by the donor star, which takes the thermal response of the outermost layers into account  when the timescale of mass transfer is so short that the relevant near-surface layers can no longer radiate away significant entropy before being lost. We find that this third criterion most consistently corresponds to the boundary between the stable, self-regulated mass transfer and unstable, runaway mass transfer shown by our mass-transfer simulations. For many cases of mass transfer from red giant donors where either the evolution accelerates to a dynamical timescale or the donor star swells far enough beyond its Roche lobe to fill its OL, or both, we find that the mass transfer can still be self-regulating and stable in our calculations. Rather than robust criteria for unstable mass transfer, these latter two should instead be employed to estimate where 1D stellar evolution codes might become unreliable in this context. 

Therefore, our results should be interpreted with care. It is safe to assume that systems that can be classified as stable by all three criteria we employed will avoid a CE due to the response of the donor star. Similarly, binaries with $q<q_{\rm qad}$ can surely be expected to encounter unstable mass transfer, leading to a CE. However, for binaries with $q_{\rm qad} < q < \{q_{\rm dyn},q_{\rm OLOF}\}$, the outcome is still uncertain, owing to the limitations of our simulations. These include the fact that we did not evolve the accretors and only considered conservative mass transfer using the 1D MESA code.

In summary, our results confirm those of previous studies: typical assumptions of polytropic donor star structure and an adiabatic response of the donor star, on which prescriptions used in many rapid BPS models are based, severely underestimate the stability of mass transfer from giant donors. Additionally, compared to studies that assume the donor always responds to the loss of mass fully adiabatically, we find mass transfer to be more stable. Because we do not impose an adiabatic response of the donor star, the sub-surface layers of giants with convective envelopes are allowed to respond to mass loss on their short thermal timescales. This postpones a fully adiabatic response and allows stable mass transfer up to a critical mass-transfer rate, which is several orders of magnitude higher than the global KH rate.

We compared our theoretical limits for stable mass transfer to the \cite{Zor11} sample of PCEBs with inferred pre-RLOF properties, and we find them to be in good agreement. Our results have important consequences for interacting binaries. Typically, the critical mass ratio we find is significantly lower than currently assumed in population calculations, meaning an increase in the number of systems predicted to avoid a CE can be expected. Additionally, we find that the critical mass ratio decreases with donor radius along the giant branch much more strongly than in the typical assumptions made in rapid BPS codes. This would change the properties of post-mass-transfer populations in a rather complicated manner that is important to explore in more detail in the future.

\begin{acknowledgements}
      The authors express gratitude to the anonymous referee, the language editor and the editor, whose comments helped improve the quality of this work.
      KDT furthermore sincerely thanks Jakub Klencki for many informative discussions and valuable advice regarding proper MESA usage and binary stellar simulations in general, and Evan Bauer for much-appreciated help understanding MESA logs and fixing errors. Lastly, KDT thanks Rob Farmer, Glenn Oomen, Ylva G\"otberg and Mathieu Renzo for advising on more specific issues. Parsing of the data from MESA simulations was done using Rob Farmer's \texttt{mesadata} module. KDT acknowledges support from NOVA.
      SJ acknowledges funding from the Netherlands Organisation for Scientific Research (NWO), as part of the Vidi research program BinWaves (project number 639.042.728, PI: de Mink).
      AGI acknowledges support from the Netherlands Organisation for Scientific Research (NWO).
      This research made use of the following \texttt{Python} (Python Software Foundation, \href{https://www.python.org/}{https://www.python.org/}) software packages and tools: \texttt{matplotlib} \citep{Hun07}, \texttt{SciPy} \citep{Vir20}, \texttt{IPython} \citep{Per07}, \texttt{NumPy} \citep{Walt11} and \texttt{Pandas} \citep{McK10}. The software references above were gathered from the Astronomy Acknowledgement Generator (\href{http://astrofrog.github.io/acknowledgment-generator/}{http://astrofrog.github.io/acknowledgment-generator/}).
\end{acknowledgements}

\bibliographystyle{aa}
\bibliography{bibtex_mt}

%
%

\clearpage
\onecolumn

\begin{appendix} 

\section{Convergence study}
\label{sec:convergence_study}

In order to make sure that our results are numerically converged, we re-ran a selection of simulations of both stable and unstable mass transfer, varying the spatial resolution of our calculations as well as the constraints on the maximum time step size systematically.

More specifically, to assess dependence of our results on spatial resolution, we increase the value of the \texttt{log\_tau\_function\_weight} parameter from 10 to 100 (we use a value of 50 in our grid calculations). This is a way to assign `meshing weight' to regions in the star based on the logarithm of the optical depth. Qualitatively, this leads to smaller allowed jumps in mesh functions and hence higher resolution in the outermost layers of the star. Whilst increasing the number of mesh-points, especially in the outermost layers, generally increases the numerical stability and accuracy of the calculations, increasing them beyond our default leads to impractically long computation times without changing our conclusions about the stability of the mass transfer or results regarding its detailed evolution.

To test the dependence of our results on the size of the time-steps, we vary the \texttt{fr} (our default is 0.01) and \texttt{delta\_lg\_star\_mass\_limit} (our default is $10^{-4}$) parameters in the ranges 0.001-0.1 and $5\cdot10^{-5}-5\cdot10^{-3}$, respectively, and adjust their associated `limit' parameters accordingly. These parameters control the maximum allowed change in the donor star's relative overflow past its Roche lobe and donor mass per time-step. We find that varying these settings has minimal effects on the simulated binaries' evolution, besides affecting the calculation time. Another related setting we varied is the \texttt{fr\_dt\_limit} option. This parameter is the lower limit (in years) to the timestep size based on the donor star's relative overflow past its Roche lobe, and thus limits the effectiveness of the parameters mentioned before. We varied this parameter between 100 and 0.01 (the MESA default is 10). We found that decreasing \texttt{fr\_dt\_limit} below its default could lead to increased numerical convergence and stability, reflected mostly in the number of time steps required to simulate challenging parts of the mass transfer as well as the number of retries and solver iterations. However, this behaviour was not monotonic and highly case-dependent. Typically, a value of \texttt{fr\_dt\_limit} = 1 provided the best balance between numerical stability and practical running times. Hence, this is our default value.

Furthermore, we experimented with two different expressions of the energy equation in the stellar structure and evolution equations \citep[see][Sect. 8 for all built-in options in MESA]{Pax18}. We found that there is no discernible difference in the evolution of the mass-transfer rate $\dot{M}_{\rm d}(t)$ between the two (\texttt{use\_dedt\_form\_of\_energy\_eqn} equal to \texttt{.true.} or \texttt{.false.}) for binaries that have mass ratios well above or below $q_{\rm qad}$. Hence, for these systems, our conclusions are not affected by the choice of energy equations. For systems nearer the boundary between stable and unstable mass transfer, we often found it necessary to `hand-pick' the most appropriate expression of the energy equation to ensure numerical stability and physical accuracy. In our grid of calculations, it did not appear that one of the two expressions we considered was consistently more appropriate than the other, and many times variation or use of other settings was necessary to ensure convergence.

\section{Supplementary tables}
\label{sec:appendix_tables}

Here we provide detailed tabulations of our main results. Table \ref{tab:mdot_results} lists for all our binary models (except for the bisection models) key aspects of the mass-transfer-rate evolution, which can prove useful as a benchmark for rapid BPS models. Table \ref{tab:main_results} enumerates for each of our grid points (the open circles in Fig. \ref{fig:model_selection}) the critical mass ratios based on the three criteria described in Sect. \ref{sec:criteria} as well as the critical thermal mass-transfer rate $\dot{M}_{\rm th, crit}$ at the start of mass transfer. Specific values of and uncertainties on $q_{\rm qad}$ were obtained by performing bisecting MESA calculations until the uncertainty decreased below 5 per cent. Grid points with larger relative uncertainty correspond to models where the donor star expanded past the size of the binary orbit, which were not bisected farther.

\begin{table}[h!]
\caption{Key aspects of the mass-transfer evolution in our binary models.}             
\label{tab:mdot_results}      
\centering          
\begin{tabular}{cccc|rr}
    \hline\hline
    $M_{\rm d}/M_\odot$ & model & $\log\, (R_{\rm d}/R_\odot)$ & $M_{\rm a}/M_{\rm d}$ & $\log\max(\dot{M}_{\rm d})$ & $\log\left\langle\dot{M}_{\rm d}\right\rangle_{\rm M}$ \\\hline
    1.0 & 0 & 0.129 & 1.2 & -9.422 & -9.58 \\
    1.0 & 0 & 0.129 & 1.1 & -9.466 & -9.61 \\
    1.0 & 0 & 0.129 & 1.0 & -9.517 & -9.647 \\
    1.0 & 0 & 0.129 & 0.9 & -9.57 & -9.69 \\
    1.0 & 0 & 0.129 & 0.8 & -9.414 & -9.724 \\
\end{tabular}
\tablefoot{The table lists for each ($M_{\rm d}/M_\odot$ , $\log\, (R_{\rm d}/R_\odot)$, $M_{\rm a}/M_{\rm d}$) in our grid the peak mass-transfer rate $\max(\dot{M}_{\rm d})$ and the mass-averaged mass-transfer rate $\left\langle\dot{M}_{\rm d}\right\rangle_{\rm M}$. For brevity, only the first five rows are shown here. The full table is available on request or in electronic form at the CDS via anonymous ftp to cdsarc.u-strasbg.fr (130.79.128.5) or via \href{http://cdsarc.u-strasbg.fr/viz-bin/qcat?J/A+A/}{http://cdsarc.u-strasbg.fr/viz-bin/qcat?J/A+A/}}
\end{table}

\begin{table}[h!]
\caption{Tabulation of our main results.}             
\label{tab:main_results}      
\centering          
\begin{tabular}{ccccccccc}
    \hline\hline
    ZAMS mass & model & $\log(R)$ & mass$\left.\right|_{\rm RLOF}$& $\log(\dot{M}_{\rm KH})$ & $\log(\dot{M}_{\rm th, crit})$ & $q_{\rm qad}$ & $q_{\rm OLOF}$ & $q_{\rm dyn}$\\
$M_\odot$ & - & $R_\odot$ & $M_\odot$& $M_\odot \text{yr}^{-1}$ & $M_\odot \text{yr}^{-1}$ & $M_{\rm a}/ M_{\rm d}$ & $M_{\rm a}/ M_{\rm d}$ & $M_{\rm a}/ M_{\rm d}$\\
\hline
    1.0 & 0 & 0.129 & 0.9995 & -6.820 & -3.990 & 0.431 $\pm$ 0.006 & $\dagger$ & $\dagger$ \\
1.0 & 1 & 0.252 & 0.9994 & -6.600 & -3.860 & 0.744 $\pm$ 0.006 & $\dagger$ & $\dagger$ \\
1.0 & 2 & 0.447 & 0.9993 & -6.163 & -3.590 & 1.031 $\pm$ 0.006 & $\dagger$ & $\dagger$ \\
1.0 & 3 & 0.645 & 0.9991 & -5.607 & -3.217 & 1.006 $\pm$ 0.006 & $\dagger$ & $\dagger$ \\
1.0 & 4 & 0.841 & 0.9988 & -5.073 & -2.865 & 0.956 $\pm$ 0.006 & $\dagger$ & $\dagger$ \\
\end{tabular}
\tablefoot{This table shows the critical mass ratios corresponding to the criteria described in Sect. \ref{sec:criteria}. For $q_{\rm dyn}$, $A_{\rm dyn} = 0.05$ was assumed. For mass ratios smaller than the tabulated values, mass transfer is unstable according to the related criterion. A dagger symbol ($\dagger$) indicates where the mass transfer became unstable and our simulations ended before reaching either the dynamical-timescale or OLOF criteria. Additionally, this table enumerates, for each single star model used as starting point for the donor star in our binary calculations, the critical mass-transfer rate and the KH mass-transfer rate (as defined in Sect. \ref{sec:crit_th}). For brevity, only the first five rows are shown here. The full is available on request or in electronic form at the CDS via anonymous ftp to cdsarc.u-strasbg.fr (130.79.128.5) or via \href{http://cdsarc.u-strasbg.fr/viz-bin/qcat?J/A+A/}{http://cdsarc.u-strasbg.fr/viz-bin/qcat?J/A+A/}}
\end{table}

\end{appendix}

\end{document}